\documentclass[a4paper,aps]{revtex4}
\usepackage[]{latexsym}
\usepackage{bm}
\usepackage{subeqnarray}
\usepackage{graphicx,amsfonts,amssymb,amsmath,amsthm,amscd}
\usepackage[english]{babel}
\usepackage{hyperref}

\theoremstyle{definition}
\newtheorem{definition}{Definition}
\theoremstyle{definition}
\newtheorem{remark}{Remark}[definition]
\newcommand{\bde}{\begin{definition}}\newcommand{\ede}{\end{definition}}
\newcommand{\bre}{\begin{remark}}\newcommand{\ere}{\end{remark}}
\newcommand{\be}{\begin{equation}}\newcommand{\ee}{\end{equation}}
\newcommand{\bea}{\begin{eqnarray}}\newcommand{\eea}{\end{eqnarray}}
\newcommand{\bsa}{\begin{subeqnarray}}
\newcommand{\esa}{\end{subeqnarray}}
\newcommand{\brr}{\begin{array}}\newcommand{\err}{\end{array}}
\newcommand{\bit}{\begin{itemize}}\newcommand{\eit}{\end{itemize}}
\newcommand{\ben}{\begin{enumerate}}\newcommand{\een}{\end{enumerate}}

\newcommand{\ba}{\begin{array}}
\newcommand{\ea}{\end{array}}

\def\lab{\label}

\def\lf{\left}

\def\non{\nonumber}
\def\rar{\rightarrow}
\def\ri{\right}\def\ti{\tilde}

\def\de{\delta}

\def\om{\omega}

\def\1{{_{1}}}\def\2{{_{2}}}

\def\noHe0{:\;\!\!\;\!\!:H_e(0):\;\!\!\;\!\!:}
\def\noHm0{:\;\!\!\;\!\!:H_\mu(0):\;\!\!\;\!\!:}

\def\lab{\label}

\def\lf{\left}

\def\non{\nonumber}

\def\rar{\rightarrow}
\def\ri{\right}\def\ti{\tilde}

\def\de{\delta}

\def\om{\omega}

\def\1{{_{1}}}\def\2{{_{2}}}

\usepackage{pifont}
\usepackage[normalem]{ulem}

\begin{document}


\title{Quantum Field Theory and Coalgebraic Logic in Theoretical Computer Science}

\author{Gianfranco Basti${}^{a}$}
\author{Antonio Capolupo${}^{b}$}
\author{Giuseppe Vitiello${}^{b*}$}
\affiliation{${}^{a}$Dipartimento di Filosofia, Universit\`a Lateranense, Roma - 00184, Italy}
\affiliation{${}^{b}$Dipartimento di Fisica E.R.Caianiello, Universit\`a di Salerno,\\ and INFN Gruppo collegato di Salerno, Fisciano (SA) - 84084, Italy\\
${}^{*}$corresponding author: 
vitiello@sa.infn.it }


\begin{abstract}

\begin{center}\textbf{Abstract}\end{center}
In this paper we suggest that in the framework of the Category Theory it is possible to demonstrate the mathematical and logical \textit{dual equivalence} between the category of the $q$-deformed Hopf Coalgebras and the category of the $q$-deformed Hopf Algebras in QFT, interpreted as a thermal field theory. Each pair algebra-coalgebra characterizes, indeed, a QFT system and its mirroring thermal bath, respectively, so to model dissipative quantum systems persistently in far-from-equilibrium conditions, with an evident significance also for biological sciences. The $q$-deformed Hopf Coalgebras and the $q$-deformed Hopf Algebras constitute  two dual categories because characterized by the same functor $T$, related with the Bogoliubov transform, and by its contravariant application $T^{op}$, respectively. The \textit{q}-deformation parameter, indeed, is related to the Bogoliubov angle, and it is effectively a thermal parameter. Therefore, the different values of $q$ identify univocally, and then label, the vacua appearing in the foliation process of the quantum vacuum. This means that, in the framework of Universal Coalgebra, as general theory of dynamic and computing systems (``labelled state-transition systems"), the so labelled infinitely many quantum vacua can be interpreted  as the Final Coalgebra of an ``Infinite State Black-Box Machine''. All this opens the way to the possibility of designing a new class of universal quantum computing architectures based on this coalgebraic formulation of QFT, as its ability of naturally generating a Fibonacci progression demonstrates.\\


\end{abstract}

\maketitle
\section{Introduction}
\label{sec:intr}
Category Theory (CT) endows modern logic and mathematics with a universal language in many senses ``wider" than set theory. By CT indeed, it is possible to demonstrate structural similarities among theories that it is impossible to discover otherwise. In this framework, there exists a growing convergent interest in literature toward the \textit{topological interpretation} of quantum field theory (QFT), both in theoretical physics and in theoretical computer science.

In this paper we present a further exemplification of the power of CT on this regard and we discuss a possible approach to topological quantum computing, in the framework of  QFT interpreted as a thermal field theory, according to a development of the original Umezawa's Thermal Field Dynamics (TFD) (Takahashi and Umezawa, 1975; Umezawa et al., 1982; Umezawa, 1993). This development is based on the systematic use of the \textit{``q}-deformed Hopf coalgebras'' for modeling \textit{open} quantum systems, where \textit{q} is a thermal parameter related to the Bogoliubov transformation.

These coalgebras are characterized by \textit{non-commutative co-products}, since they represent states of the system and  states of its thermal bath, and these cannot be treated on the same basis and cannot be exchanged one with the other. We thus introduce the \textit{doubling of the degrees of freedom} (DDF), based on the doubling of the states of the Hilbert space, including the system states and the thermal bath states
(Blasone et al., 2011). This opens the way to a topological QFT modeling of quantum dynamics for systems \textit{far-from-equilibrium,} and not only in a \textit{near-to-equilibrium} condition.

Our  approach to a topological QFT is, therefore, a development of TFD as far as is based on the mechanism of spontaneous breakdown of symmetry (SSB) with the related Goldstone theorem, the Goldstone boson condensates (Goldstone et al., 1962) and the Nambu-Goldstone (NG) \textit{long-range correlation} modes or NG quanta, familiar in condensed matter physics and elementary particle physics.

These NG correlations allow the existence of infinitely many \textit{non-unitarily inequivalent representations} of the canonical commutation relations (CCR's) in QFT (non-unitarily inequivalent Hilbert spaces of the system states) (Blasone et al., 2011).  The DDF allows then to use the \textit{minimum free energy function and measure} as an intrinsic dynamical tool of choice among states, so to grant a dynamic determination of the orthonormal basis of the Hilbert space. It introduces at the same time the notion of the quantum vacuum (QV) or ground state \textit{foliation}, as a robust principle of ``construction'' and of ``memory'' used by nature for generating ever more complex systems.

It is therefore not casual that during the last ten years the QV-foliation in QFT has been successfully applied to solve dynamically the capacity problem of the \textit{long-term memories}  -- namely, the ``deep beliefs'' in the computer science jargon -- in the living brain, interpreted as a ``dissipative brain'', i.e.,``entangled'' with its environment (thermal bath) via the DDF formalism (Vitiello, 1995, 2001, 2015; Freeman and Vitiello, 2006, 2008; Capolupo et.al., 2013;  Basti, 2013).

These neurophysiological studies suggest that the DDF based on a coalgebraic modeling of open systems in QFT could be, in computer science, an effective possible solution of the problem, so-called of  ``deep learning'', arising with ``big-data'' modeling and, overall, in dealing with an effective  computational management of (infinite) ``streams'' (think, for instance, at the internet streams) characterized  by a continuous change of the ``hidden'' higher-order correlations among data, which are absolutely unpredictable by the classical statistical tools. In this case, the DDF formalism could endow a topological quantum computing system with a dynamic (= automatic) rearrangement of the dimensions (degrees of freedom) of the representation space of the system, so ``to lock it dynamically'' on the ever changing degrees of freedom of the stream (interpreted as the system thermal bath), by using the minimum free energy function as a truth evaluation function (Basti, 2017).

Effectively, one of the pillars of the topological quantum computing, i.e., the fundamental \textit{Stone representation theorem for Boolean algebras} (Stone, 1936), is strictly related with the topological interpretation of QFT. Stone, indeed, arrived at this theorem after a deep reflection on Hilbert spectra that leaded him to demonstrate five years before in 1931, with J. von Neumann, a fundamental theorem at the basis of the formalism for QM we discuss below (von Neumann, 1955). The topologies of the Stone spaces are  effectively the same topologies of  the \textit{C*}-algebras, and are associated with Hilbert spaces via the GNS-construction (Landsman 2011); see also Appendix \ref{dual_alg_coalg})\footnote{It is useful to recall that the topological spaces (and then the Stone-spaces) related to the \textit{C*-}algebras are \textit{non-commutative} and then dual to \textit{C*-}algebras that are \textit{commutative}}. The Stone representation theorem for Boolean algebras demonstrated that each Boolean algebra is isomorphic with a partially ordered set (see Appendix \ref{sect:app}) of clopen sets (closed-open sets, effectively intervals of reals) on a Stone-space. Moreover, the category of Boolean algebras and the category of the Stone-spaces are \textit{dual} to each other, in the sense that, given two Boolean algebras \textit{A, B}, and the respective Stone-spaces $S(A), S(B)$,  to a continuous function from \(S(A)\) to \(S(B)\) it corresponds a monotone function from \textit{B} to \textit{A }.

Afterwards, a seminal work of S. Abramsky at the end of the 80's (Abramsky, 2005), demonstrated the \textit{dual equivalence} (see Appendices \ref{dual} and \ref{dual_alg_coalg}) between the category of coalgebras defined on Stone-spaces, and the category of Boolean algebras, so to introduce the possibility of defining coalgebraically, on the physical states of the system, the semantics of computing systems (Venema, 2007).

For sake of completeness, it is fundamental to recall that the momentous result of Abramsky depends on modeling coalgebras over P. Aczel's ``non-well founded (NWF) sets" (Aczel, 1988), that is, a set theory where the ``regularity axiom" of the ZF (Zermelo-Fraenkel)  set theory does not hold, and then no set ``total ordering" (see Appendix \ref{sect:app}) is allowed. In fact, because of Aczel's ``anti-foundation axiom", the set self-inclusion is possible, and then ``unbounded chains of set inclusion" are allowed. NWF sets are then what is necessary for a coalgebraic modeling of infinite streams (see Sect. \ref{ssect:qv}). Indeed, in NWF-set theory the powerful ``final coalgebra theorem" holds
(Aczel and Mendel, 1989). By NWF set theory it is possible, therefore, to justify formally the coalgebraic notions of: 1) (upper bounded) \textit{coinduction}, as method of set definition and proof, and dual to the algebraic notion of (lower bounded) \textit{induction} as method of set definition and proof; as well as of 2) \textit{bisimulation} (and \textit{observational equivalence}: see Sect.\ref{ssect:qv} for an application), as dual to the algebraic notion of \textit{congruence}.

In this way, Abramsky's construction can be read as the \textit{dual equivalence} between an \textit{initial} algebra (with a least fixed point), and a \textit{final} coalgebra (with a last fixed point), so to make upper and lower \textit{bounded} the Boolean algebra operators (Rutten, 2000; Sangiorgi and Rutten, 2012), and solving coalgebraically a millennial (since the time of Eudoxus' ``exhaustion method" of demonstration) problem of any \textit{constructive} logic and mathematics (Nelson, 1986; Perrone, 1995). Namely, the lacking of ``a least upper bound" of any inductive construction by recursiveness, for which it becomes necessary to refer to the \textit{infinitistic } notions of higher order logic and mathematics (like the Platonic ``circle" as existing necessary limit of Eudoxus' construction). By using a paraphrase  of Abramsky's words, what is necessary for computations are, neither ``finitistic", nor ``infinitistic" notions, but \textit{finitary} ones, intended as ``dual limits/colimits of finite successions" (Abramsky, 2005).

As we see in this paper, this coalgebraic construction can be extended also to the duality between the categories of \textit{q-}deformed Hopf coalgebras and algebras characterizing the thermal QFT, and then extended to the DDF principle of QFT dissipative systems (see Appendix \ref{dual_alg_coalg}). Indeed, the Gelfand-Naimark-Segal (GNS) construction in the \textit{C*}-algebra formalism (Emch, 1972; Bratteli and Robinson, 1979) is compatible with the DDF formalism (Ojima, 1981).

Therefore, our general scheme of coalgebraic modeling of dissipative QFT system shares the same topology of the \textit{C*}-algebras, and then of the Stone-spaces. Moreover, because of the strict relationship between the implementation of a two-state qubit in our topological quantum computing structure, and the measures of free energy minimum naturally deriving from the equations, the qubit associated with the DDF has an intrinsic \textit{semantic} value of a ``truth"-``falsehood" Boolean operator (Basti, 2017). Significantly, the iterated application of such a ``semantic qubit" is able to construct Fibonacci's progressions, manifesting itself as a ``universal computing machine" of the coalgebraic type (see Sects. \ref{ssec:qbit2} and \ref{ssect:qv})

The plan of the paper is therefore the following. We shortly discuss in Section II the notions of topological quantum computing, and of quantum thermal systems, both in the classical statistical mechanics interpretation of topological QFT.  Section III is devoted to the discussion of $q$-deformed Hopf coalgebras in QFT, with reference to the doubling of the degrees of freedom of the system, its role in thermal field theories also in relation with \textit{C*}-algebra formalism, as well as the implementation of a ``semantic qubit" in dissipative quantum systems. In Section IV, we complete our computational interpretation of thermal QFT in the framework of the ``Universal Coalgebra", as a general theory of dynamic and computational systems. Conclusions are presented in Section V and formal details are presented in the Appendices.

\section{Topological quantum computing in  QFT}

\subsection{Topological quantum computing}

Within the approach of statistical mechanics to many-body physics, the possibility of topological quantum computing (QC) in the framework of topological QFT (Freedman et al, 2002) depends on the possibility of implementing \textit{topological orderings} based on \textit{non-commutative} products that are essential for computations, and then it requires the use of \textit{non-Abelian algebras} (Nayak et al., 2008). The strong interest in this field is today justified by the emergence of ``exotic" quantum phases of matter characterized by \textit{topological orderings} at close to zero temperatures, i.e., by patterns of long range quantum entanglement, which manifest at the macroscopic level as robust phenomena of  ground state degeneracy -- e.g., 1D superconductivity. One of the 2D topological orderings, related with the theoretical possibility of constructing fault-tolerant quantum computers, is the so-called ``fractionary quantum Hall states". These are displayed by topological phases of matter whose excitations are neither fermions nor bosons, but quasi-particles ranging continuously between Fermi-–Dirac and Bose–-Einstein statistics that can obey a \textit{non-Abelian braiding statistics}, i.e., the so-called \textit{non-Abelian anyons\footnote{The reference to non-Abelian groups is essential for preserving the multiplication order of braiding operations, so to satisfy also in this case the systematic usage of \textit{conformal} algebras in operator algebra applied to the topological QFT modeling, i.e., the so-called ``conformal quantum field theory". The possibility of universal quantum computations based on non-Abelian statistics has been mathematically demonstrated in 2002 (e.g., the ``Fibonacci anyon model" or ``Golden theory") (Freedman, 2002). This demonstration is, indeed, at the origins of the further studies in this field.}}.  The quantum logical gates are implemented ``by braiding" quasi-particle waves, and then measuring     the multi-quasi-particles states.  The fault-tolerance of this architecture depends on the non-local encoding of the states of quasi-particles, so to make it immune to errors due to local perturbations (Nayak et al., 2008). Finally, the theorical derivation of \textit{Fibonacci's sequences} by  recursive processes in such a topological quantum computing system, demonstrates the \textit{universality} of such a ``Fibonacci's anyonic model" or ``Golden machine" (Freedman et al., 2002). Effectively, all the interest for this ``non-Abelian" quantum computing architecture originally derived from such a demonstration. The problem remains, however, if non-Abelian anyons have been effectively observed and measured, as announced by Bell Laboratories in an implementation of such a quantum computation architecture, by using three quantum Hall interferometers on a gallium arsenide substrate  at $\approx$ 0\textordmasculine K   temperatures (Willett et al., 2013; von Keyserlingk et al., 2015).

The deeper problem for the topological QFT interpretation of condensed matter physics is whether topological orderings are stable at temperatures $\gg$0\textordmasculine K, or, in other terms, whether patterns of long range correlations occur in dynamic systems far from equilibrium and endowed with several different phases. In this case, indeed, topological orderings for each different phase have to extend in some way also to the thermal bath of such systems. Solving formally this problem is essential for QFT modeling dissipative condensed matter systems (think of chemical and biological systems, for instance) as well as for constructing systems performing {\it dynamical} quantum computations. Effectively, our coalgebraic interpretation of QFT dissipative systems is able to generate Fibonacci's progressions by iterative applications of the ``semantic qubit" implemented in these systems (see Sect. \ref{ssec:qbit2}). This justifies an interest also in this type of QFT modeling of universal quantum computation. Therefore, for better appreciating the core differences between a thermal QFT based on TFD, and the conventional interpretation of quantum thermodynamic systems based on statistical mechanics, let us illustrate briefly in the following subsection B a recent result in thermal QFT theory based on the conventional approach.

\subsection{Thermal theories in  QFT}

There exists since many years a systematic effort in developing thermal theories in  QFT  by using \textit{covariant} von Neumann algebras\footnote{\label{homomorphism}That is, they are related by a \textit{covariant} homomorphism, i.e., in the CT jargon, by a \textit{functorial }mapping \textit{F} preserving all the arrow (morphism) \textit{f, g } directions and the composition orders, \(f\circ g,\) between the two homomorphic structures. The homomorphism between structures is, on the contrary, \textit{contravariant} if  it is reversing all the arrows and the composition orders. In such a case, the two structures are \textit{dually equivalent} for the contravariant application of the same endofunctor \textit{F}. It is important do not confuse homomorphism with \textit{homeomorphism}, designating an equivalence between topologies. See Appendix }. Recently,  a significant -- because model independent -- positive answer in the direction of a topological approach to non-equilibrium quantum thermal theory has been obtained by formalizing, in the framework of a bi-dimensional conformal field theory, a thermal system at
\textit{near-to-equilibrium} conditions (Holland and Longo, 2016). According to this scheme,  a non-equilibrium non-homogenous quantum system can be modeled in SM in terms of two infinite energy reservoirs initially at different temperatures and different chemical potentials, and then interacting at the boundaries through a defect line via a energy flux from one reservoir to the other. By limiting ourselves for sake of simplicity to the case without chemical potentials, the evolution in time of the total system towards \textit{a local non-equilibrium steady state}, satisfying in the infinite volume limit the statistical mechanics KMS-condition, can be represented in conformal QFT by using two nets of von Neumann algebras defined on the same Minkowski space-time $M$.

Initially, at time $t<0$ on the 
light cone $O$ of $M$, the two non-interacting sub-systems are represented as two covariant nets of von Neumann algebras\footnote{\label{conformal}They are two conformal nets defined on the Minkowski spacetime. Indeed, the M\"obius group PSL\((2,\mathbb{R})\) acts on each compactified light ray line \(\Lambda_\pm\cup\{\infty\}\) of $M$, by linear fractional transformations, hence we have a local (product) action of PSL\((2,\mathbb{R})\) $\times$ PSL\((2,\mathbb{R})\) on $M$. Therefore, a \textit{local M\"obius covariant net $\mathcal{B}$} on $M$ is a map, \(\mathcal{B}  :O \in \mathcal{K}\mapsto\mathcal{B}(O)\), where \(\mathcal{K}\) is the set of double cones of $M$, and \(\mathcal{B}(O)\)'s  are von Neumann algebras on a fixed Hilbert space $\mathcal{H}$.}, $\mathcal{B}^L/\mathcal{B}^R$, each occupying the left, \textit{L}, or the right, \textit{R}, halves, $M_L/M_R$, of $M$, and separated by an impenetrable wall. Each net of von Neumann algebras $\mathcal{B}$, is therefore insisting on its own Hilbert space $\mathcal{H}$, and it is associating the $C^*$-algebra $\mathfrak{B}$ generated by \(\mathcal{B}\) to a given region of \(M\). From now on, we use a calligraphic letter to denote a net of von Neumann algebras, and the corresponding Gothic letter to denote the associated \(C^{*}\)-algebra.

Let us consider now the dynamics of the composite system \textit{before} and \textit{after} the two systems are in contact.

\textbf{\\Before contact}.
At time $t<0$, the two systems $\mathcal{B}^L$ and $\mathcal{B}^R$ live independently in their own half plan $M^L$ and $M^R$ and their own Hilbert space. They are each at equilibrium state, at different temperatures. We can thus consider the KMS-state \(\varphi^{L/R}_{\beta_{L/R}}\)on \(\mathcal{B}^{L/R}\) with respect to the translation automorphism group $\tau$, with \(\mathcal{B}^{L}\neq \mathcal{B}^{R}\). Indeed, the composite system is described by the net on \(M_L\cup M_R\) given by
\begin{equation}
\mathcal{K}(M_L)\ni O \mapsto \mathcal{B}^L (O),\:\;\quad\mathcal{K}(M_R)\ni O \mapsto \mathcal{B}^R (O)
\end{equation}
where $\mathcal{K}$ is the set of the double light cones $O$ on $M$. The $C^*$-algebra of the composite system is \(\mathfrak{B}^L(M_L) \otimes\mathfrak{B}^R(M_R)\) and the state of the system is
\begin{equation}
\varphi=\varphi^{L}_{\mathcal{B}_L|\mathfrak{B}^L(M_L)}\otimes \varphi^{L}_{\mathcal{B}_R|\mathfrak{B}^R(M_R)}
\end{equation}
where $\varphi$ is a non-equilibrium steady state, but not a KMS state.
\\ Let us introduce now the further notations: 1) \(V_\pm = \{(t,x):\pm t>|x|\}\) as the forward/past \textit{light cone}; and 2) \(W_{L/R}=\{(t,x):\mp x>|t|\}\) as the left/right \textit{wedge} in $M$.\\

\textbf{\\ After contact.} At time \(t=0\) we put the two systems $\mathcal{B}^L$ and $\mathcal{B}^R$ in contact through a \textit{transmissive} phase boundary, and the time-axis as the defect line. In this case, $\mathcal{B}^L$ and $\mathcal{B}^R$ are nets acting on a \textit{common, ``doubled"} Hilbert space $\mathcal{H}$. By denoting with \(O_1\subset M_L, \;O_2\subset M_R\) double cones, the von Neumann algebras \(\mathcal{B}^L (O_1)\) and \(\mathcal{B}^R (O_2)\) \textit{commute} if $O_1$ and $O_2$ are space-like separated, so that also the associated $C^*$-algebras \(\mathfrak{B}_L (W_1)\) and \(\mathfrak{B}_R (W_2)\) \textit{commute} as well.

We want now to describe the state $\psi$ of the global system for time $t>0$. We consider, therefore, the von Neumann algebras of the doubled system   generated by \(\mathcal{B}^L (O)\) and \(\mathcal{B}^R (O)\): \begin{displaymath}
\mathcal{D}(O)\equiv \mathcal{B}^L(O)\vee \mathcal{B}^R(O),\quad O\in\mathcal{K}
\end{displaymath}
The origin 0 is the only $t=0$ point of the defect line, so that the observables localized in the causal complement \(W_L \cup W_R\) of the 0 (i.e., for $t<0$) are not submitted to the effect of the contact. Therefore, $\psi$ should be a natural state on $\mathfrak{D}$ that satisfies
\begin{displaymath}
\psi_{|\mathfrak{B}^L(W_L)}=\varphi^{L}_{\beta_L|\mathfrak{B}^L(W_L)},\quad \psi_{|\mathfrak{B}^R(W_R)}=\varphi^{R}_{\beta_R|\mathfrak{B}^R(W_R)}.
\end{displaymath}
That is, \(\psi\) must be a \textit{local thermal equilibrium state} on \(W_{L/R}\). However, because of the contact along the temporal defect line, \(\mathfrak{B}^L(M_L)\) and \(\mathfrak{B}^R(M_R)\) are not independent, the existence of such a state $\psi$ is not obvious at all.
Anyway, by referring to the paper here discussed for more details, the significant result reached by the Authors in the framework of the asymptotic condition of the perturbative methods, is that such an initial state $\psi$ (with impenetrable walls) approaches, via an evolution $\psi \cdot\tau_t$ to \textit{a non-equilibrium steady state} $\omega$ as $t \rar +\infty$, where $\tau_t$ is a time translational automorphism.
\\
For obtaining such a result, it is sufficient to demonstrate that \(\psi_{|\mathcal{D}(O)}= \omega _{\mathcal{D}(O)}\)\ \textit{if} \(O \in \mathcal{K}(V_+)\). Thereafter, let $O \in \mathcal{K}(M)$, and $Z \in \mathcal{D}(O),$ be any local observable of the system. If the time translated double cone $O_t\equiv O+(t,0)$ enters and remains in $V_+$ for \textit{t} larger than a given time $t_O$, then  \textit{for every} $Z \in \mathfrak{D}$, we have\footnote{In fact, for the $GNS$-construction, the limit holds true if $Z$ belongs to the norm dense subalgebra of the $C^*$-algebra \(\mathfrak{D}\) generated by the \(\mathcal{D}(O)\)'s, \(O \in \mathcal{K}\), and then for all \(Z \in \mathfrak{D}\) by density approximation.}
\begin{equation}
\lim_{t \to \infty}\psi(\tau_t(Z))= \omega(Z).
\end{equation}
This result is relevant because the non-equilibrium steady-state $\omega$ is determined only by the temperature of the reservoirs, and by the nature of the transmissive phase boundaries.

The exclusive \textit{commutative} character of the von Neumann and of the $C^*$-algebras involved, explains why a topological quantum computing on this basis needs ``exotic" phases of matter for implementing topological non-commutative orderings. Finally, the open question remains whether, in a general covariant formalism in quantum thermodynamics, we have to consider all the states on the same ground, or there exists some maximal entropy principle (minimum free energy) able to select among states (Connes and Rovelli, 1994).

\section{$Q$-deformed Hopf coalgebras in QFT}
\label{sec:qdef}
\subsection{The doubling of the degrees of freedom in QFT\textit{}}
\label{ssec:doubl}

The conventional approach in introducing the QFT algebraic structure consists in assigning the canonical commutator or anticommutator relations for the boson or fermion case, respectively. However, one needs also to specify which one is the prescription for adding primitive observables such as energy, angular momentum, etc. It could seem that such a prescription does not belong to the algebraic structure of the theory. Effectively, this is not the case. In fact, in order to specify, e.g., the total energy $E$ of two identical particles, one writes $E = E_1 + E_2$. The meaning of the labels in such a formula is that $E_1$ refers to the first particle, and $E_2$ to the second particle. However, it is easy  to realize that $E_1=E \times 1$ where the index 1 thus refers to the first position. Similarly,   $E_2 = 1 \times E$. Thus, $E= E_1 + E_2 = E \times 1 + 1 \times E$, and similarly   $J = J_1 + J_2 = J \times 1 + 1 \times J$, for the angular momentum, which are nothing but the  \textit{commutative coproducts} of a coalgebra. Here, ``commutative" refers to invariance of the coproduct under the permutation $1\leftrightarrow2$, as it needs to be on the premise that the particles are identical. We should be able therefore to go from the algebra $\cal A$ for the single particle to the algebra for two of them, namely, ${\cal A} \rar {\cal A} \times {\cal A}$. Of course, we need also to be able to go back to a single particle, namely, ${\cal A} \times {\cal A}\rar{\cal A}$.

The conclusion is that the basic algebra to start with in QFT is a \textit{bialgebra}, that is the \textit{Hopf algebra} (Blasone et al., 2011; Celeghini et al., 1998).
We thus see that the ``doubling of the degrees of freedom'' (DDF) implied in the Hopf mapping ${\cal A} \rar {\cal A} \times {\cal A}$ arises as a natural requirement in setting up the QFT algebraic structure.

Most interesting is the case when the two systems need to be treated not on the same footing, as, for example, in thermal field theory, or when dealing with open systems in general, where the system under study and its thermal bath or environment are not exchangeable. In these cases the proper tool is provided by the $q$-deformed Hopf algebras with \textit{non-commutative} coproducts, e.g., $\Delta a_{q} = a \times q + q^{-1} \times a \equiv a \,q \,+ \, q^{-1}\, {\tilde a}$, with $a \in {\cal A}$, and its hermitian conjugate, $\Delta a^{\dag}_{q} = a^{\dag} \times q^{*} + (q^{-1})^{*} \times a^{\dag} $. Of course if $q$ is real, $q^* = q$; if it is imaginary, then $q^* = q^{-1}$. The deformation parameter $q$ may depend on temperature, decay constants, etc.. The QFT formalism of the DDF has been introduced and used in many applications of the TFD formalism for many-body systems (Takahashi and Umezawa, 1975; Umezawa et al., 1982; Umezawa, 1993; Blasone et al., 2011).

The Hopf coalgebra thus describes the {\it doubling} of the degrees of freedom $a \rightarrow \,\{a,  \,  {\tilde a}\}$ and of the state space ${\cal F} \rightarrow {\cal F} \times {\tilde{\cal F}}$, with the operators $a$ and  ${\tilde a}$ acting on ${\cal F}$ and ${\tilde{\cal F}}$, respectively.  We stress that the associated Hopf algebra is, as said, a non-commutative coalgebra.

As far as no misunderstanding occurs, for simplicity we omit subscripts denoting specific properties of the operators $a$ (e.g. their quantum numbers as the dependence from the momentum $\mathbf k$,  $a_{\mathbf k}$).
The Hopf non-commutative structure determines in a crucial way the very same structure of the space of the states.
For our discussion in connection with coalgebraic logic and computer science, it is interesting to see in some details how this happens.

To this aim we observe that one can work in the hyperbolic function basis $\{e^{+\theta}, \, e^{-\theta} \}$, or in the circular function basis  $\{e^{+i\, \theta}, \, e^{- i\, \theta} \}$. As well known (Blasone et al., 2011), physically this corresponds to work with bosons or fermions, respectively, as implied by algebraic constraints on which here for brevity we do not insist. To be concrete, in the following we work with real $q(\theta) = e^{\pm\,\theta}$, i.e., with bosons, also because the most effective implementation of this architecture is in quantum photonics (see Conclusions).  By proper algebraic manipulations and linear combinations of the $\Delta a_{q(\theta)}$  one can then obtain (see (Celeghini et al., 1998))  the operators $A(\theta), \, {\tilde A}(\theta)$:
\bea \label{bogol1}
A(\theta) &=& A \cosh \, \theta   - {\tilde A}^{\dag} \sinh \, \theta , \\ \label{bogol2}
{\tilde A}(\theta) &=& {\tilde A} \cosh \, \theta   -  A^{\dag} \sinh \, \theta .
\eea
The canonical commutation relations are
\bea \label{ccr}
[A(\theta), A(\theta)^{\dag}] = 1 , \qquad  [{\tilde A}(\theta), {\tilde A}(\theta)^{\dag}] = 1 ,
\eea
All other commutators equal to zero. The Eqs.~(\ref{bogol1}) and (\ref{bogol2}) are nothing but the Bogoliubov transformations for the $\{A,  \,  {\tilde A}\}$ couple. They provide an explicit realization of the mapping of the $q-$deformed Hopf coalgebras, ${\cal A} \rar {\cal A} \times {\cal A}$.

The key point in considering the Bogoliubov transformations is that they reveal the main distinguishing characteristics of QFT, namely the existence of infinitely many {\it unitarily inequivalent}, and therefore physically inequivalent, representations of the CCR's, each one labelled by one of the specific values taken by the
$q(\theta)$-deformation parameter under the variations of $\theta$ in its definition range. We have, therefore, the $q(\theta)$-{\it foliation} of the state space. We recall that due to the Stone-von-Neumann theorem (von Neumann, 1955) in QM all the representations of the CCR are unitarily equivalent. In QFT each representation describes a specific phase of the system. Phase transition occurs when one moves through physically different phases (e.g. the ferromagnetic and the non-ferromagnetic phase; the superconductive and the normal (non-superconductive) phase, etc.).

Let $|0\rangle \equiv |0\rangle \times |0\rangle$ denote the vacuum annihilated by $A$ and $\tilde A$: $A |0\rangle = 0 = {\tilde A} |0\rangle$. $|0\rangle$ is not   annihilated by ${A}(\theta)$ and ${\tilde A}(\theta) $ (Eqs.~(\ref{bogol1}) and (\ref{bogol2})). By restoring the subscript $\bf k$, the vacuum annihilated by these last operators is
\be \lab{vactheta}
|0 (\theta)\rangle_{\cal N} = e^{i \sum_{\mathbf k} \theta_{\kappa} G_{\mathbf k}}|0\rangle =
\prod_{\mathbf k} \frac{1}{\cosh \, \theta_{\kappa}} \exp(\tanh \theta_{\kappa} A^{\dag}_{\kappa}{\tilde A}^{\dag}_{\kappa} )\, |0\rangle ,
\ee
where the meaning of the subscript $\cal N$ will be clarified in the following, $\theta$ denotes the set $\{\theta_{\kappa}, \forall {\bf k}  \}$ and ${}_{\cal N}\langle 0(\theta)|0(\theta) \rangle_{\cal N} = 1$. $G_{\mathbf k} \equiv -i\,(A^{\dag}_{\mathbf k}{\tilde A}^{\dag}_{\mathbf k} \, - \,A_{\mathbf k}{\tilde A}_{\mathbf k} )$ is the generator of the Bogoliubov transformations (\ref{bogol1}) and (\ref{bogol2})\footnote{This is also related with the squeezed coherent state generator in quantum optics,  and with quantum
Brownian motion (Blasone et al., 1998).}.
Eq. (\ref{vactheta}) shows that infinitely many couples $A^{\dag}_{\mathbf k}{\tilde  A}^{\dag}_{\mathbf k}$ are {\it condensed} in the ``vacuum"
$|0(\theta) \rangle_{\cal N}$.

By using the formula $\sum_{\mathbf k} = (V/(2 \pi)^3) \int d{\bf k}$, one realizes that $\exp \, i \sum_{\mathbf k} \theta_{\kappa} G_{\mathbf k}$ diverges in the infinite volume limit $V \rightarrow \infty$, so that in such a limit $\langle 0|0(\theta) \rangle_{\cal N}  \rightarrow 0$ and ${}_{\cal N} \langle 0(\theta')|0(\theta) \rangle_{\cal N}  \rightarrow 0$, $\forall \theta \neq \theta'$  . This shows that as $V \rightarrow \infty$ the state space splits in infinitely many inequivalent representations, each labelled by a specific $\theta$-set $\{ \theta_{\mathbf k} = \ln \, q_{\mathbf k}, \, \forall {\bf k} \}$ (foliation).

One can also show that $|0 (\theta)\rangle_{\cal N} $ is a generalized $SU(1,1)$ coherent state of condensed couples of  $A$ and $\tilde A$ modes.
These are entangled modes in the infinite volume limit. Coherence and entanglement are thus  dynamical consequences of the deformed Hopf coalgebras. Note that the generator $G_{\mathbf k}$ is essential part of the system Hamiltonian.
It can also be shown that the set of all the unitarily inequivalent representations $\{ |0 (\theta)\rangle_{\cal N}, \forall \theta \}$ is a K\"ahlerian manifold, which is a symplectic one, and trajectories in such a manifold can be shown to be classical chaotic trajectories (Vitiello, 2004).

\subsection{The doubling formalism and the Gelfand-Naimark-Segal construction in the $C^{*}$-algebra formalism}
\label{ssec:Cstar}

The above discussed $q$-deformed Hopf algebraic structure of QFT is intimately related to the $C^{*}$-algebra formalism. This establishes in an explicit formal way the link between Stone topologies, the topologies of $C^*$-algebras and the DDF in our QFT scheme. Aimed to such a goal, we observe that in the statistical mechanics formalism of Haag-Hugenholtz-Winnik (HHW) (Haag et al., 1967; Kubo, 1957; Martin and Schwinger, 1959) at the thermodynamical equilibrium the Gibbs state given by
\be \label{gibbs}
\om ({\cal O}) \equiv \frac{Tr (e^{-\beta H} {\cal O})}{Tr (e^{-\beta H})}= \langle {\cal O} \rangle
\ee
is characterized by: i) the KMS (Kubo-Martin-Schwinger) condition (Kubo, 1957; Martin and Schwinger, 1959)) stating that for any two operators ${\cal O}$ and ${\cal P}$ of the system under study, belonging to the operator $C^{*}$-algebra, 
there exists a function $F_{{\cal O}{\cal P}}(z)$, with $z$ a complex number, holomorphic in $0 < Im z <  \beta$,  continuous in $0 \leq Im z \leq \beta$ and satisfying the relations
\be \label{KMS}
F_{{\cal O}{\cal P}}(t) = \om ({\cal O}\alpha_{t} ({\cal P})),   \qquad F_{{\cal O}{\cal P}}(t + i \beta) = \om (\alpha_{t} ({\cal P}){\cal O})   \quad {\rm with}~~ \alpha_{t}({\cal P})= e^{i Ht} {\cal P}e^{-i Ht},  \quad {\rm for} ~~ t \in R
\ee
and ii)   $\om ({\cal O}^{\dag}{\cal O}) = 0 ~$  for  ${\cal O} \in {\cal M} ~~~$ $\Rightarrow {\cal O} = 0$, with ${\cal M}$ the system operators algebra. Note that the Gibbs state is a mixed state, not a pure state. The KMS condition can be written as $\langle {\cal O}{\cal P} (t)\rangle = \langle {\cal P} (t -i \beta) {\cal O} \rangle$.

A key relation is established between TFD and the Gibbs states by noticing that for any operator ${\cal O}$ it is $\langle 0(\theta)| {\cal O}|0(\theta) \rangle  = \langle {\cal O} \rangle = \om ({\cal O})$, where Eq.~(\ref{gibbs}) has been used for the last equality. Obtaining the first equality is the starting point in the TFD construction, as we will see below (cf. e.g. Eq.~(\ref{tracetherm})). In establishing the relation between TFD and the HHW formalism (and the Tomita-Takesaki formalism (Takesaki, M., 1970; Emch, 1972; Bratteli and Robinson, 1979) a central role is played by the modular conjugation operator $J$ and the modular operator $\exp({-\beta {\overline{H}}})$, with ${\overline{H}} \equiv H - {\tilde H}$, $~H = E \, A^{\dag}A$, $~{\tilde H}= E \, A^{\dag}A$, defined by
\be \label{Oj}
J\,\exp({-\frac{\beta {\overline{H}}}{2}}) M |0(\theta) \rangle = M^{\dag} |0(\theta) \rangle,     \quad     {\rm for} ~~~~ M \in {\cal M},
\ee
where $M$ represents $A$ or $A^{\dag}$. One can show that ${\overline{H}}|0(\theta) \rangle =0$ and $\exp({i\overline{H}t}){\cal M}\exp({-i\overline{H}t}) = {\cal M}$.   $J$ is an antiunitary operator satisfying
\be \label{Oj2}
J^{2} = 1, \qquad J \,|0(\theta) \rangle = |0(\theta) \rangle,
\ee
and $J{\cal M}J = {\cal M}'$, $~~JMJ = {\tilde M}$, $~~J\overline{H}J = - \overline{H}$, $~~\overline{H}= H - JHJ$. $~~{\cal M}'$ is called the commutant of ${\cal M}$. One can show (Ojima, 1981) that the KMS condition can be derived from Eq.~(\ref{Oj}), which in turn is central in the relation between TFD and HHW. This leads us to the conclusion that, as observed above, the doubling formalism can be implemented by the GNS construction in the $C^{*}$-algebra formalism of statistical mechanics.

\subsection{The Bogoliubov transformations and the $q$-deformed Hopf coalgebras}
\label{ssec:bogol}

There is one more aspect of QFT and its intrinsic algebraic Hopf structure which is relevant from the perspective of the coalgebraic logic and computer science:
the tilde modes provide the {\it intrinsic dynamic (coalgebraic) reference} (semantics) for the
non-tilde modes. The coalgebra structure of the doubling both the space, and the operators, turns into a strict
correspondence between each operator and its tilde-copy (the doubled operator) so that one of the two provides the {\it address} of the other one. The result is the self-consistent {\it dynamical} inclusion of the ``reference term'' in the logical scheme. A sort of {\it contextual self-embedding}, or {\it dynamical generation of meaning}, a ``local'', not ``absolute'', but crucially meaningful \textit{truth,} singled out of the infinitely many possibilities offered by the infinitely many representations of the CCRs.

The simplest example is perhaps obtained by explicitly computing  the expectation value of the number operator $N_{A_{\bf k}} = A^{\dag}_{\mathbf k} A_{\mathbf k}$ in the ground state $|0(\theta) \rangle_{\cal N} $. By inverting the Bogoliubov transformations,
we have $\forall {\bf k}$
\be \label{opnum}
{\cal N}_{A_{\bf k}}(\theta) \equiv {}_{\cal N} \langle 0(\theta)| A^{\dag}_{\mathbf k} A_{\mathbf k}|0(\theta) \rangle_{\cal N}  =
{}_{\cal N} \langle 0(\theta)|{\tilde A}_{\mathbf k}(\theta) {\tilde A}^{\dag}_{\mathbf k}(\theta)|0(\theta) \rangle_{\cal N}  \,  = \sinh^{2} \, \theta_{\mathbf k},
\ee
which shows that for any $\bf k$ the only non-vanishing contribution to the number of non-tilde modes ${\cal N}_{A_{\bf k}}(\theta)$  comes from the tilde operators. In this sense, these last ones constitute the dynamic {\it address} for the non-tilde modes. Through them the total number ${\cal N}_{A_{\bf k}}(\theta)$   of the $A_{\bf k}$ condensate is determined. Of course, the reverse is also true, namely the only non-zero contribution to ${\cal N}_{{\tilde A}_{\bf k}}(\theta)$ comes from the non-tilde operators, and ${\cal N}_{A_{\bf k}}(\theta) - {\cal N}_{{\tilde A}_{\bf k}}(\theta) = 0$.

The whole condensate content of
$|0(\theta) \rangle_{\cal N} $ is thus specified by the $\cal N$-set $\equiv \{ {\cal N}_{A_{\bf k}}(\theta), {\cal N}_{A_{\bf k}}(\theta) = {\cal N}_{{\tilde A}_{\bf k}}(\theta), \forall {\bf k}  \}$. Such a $\cal N$-set is called the {\it order parameter}. It provides a characterizing parameter for the $\theta$-vacuum
$|0(\theta) \rangle_{\cal N} $ and explains the meaning of the $\cal N$ subscript. Its knowledge constitute the ``end point'' of the computation, the searched {\it result}.

We thus stress once more the self-consistency of the QFT construction, by which the address or reference to the $A$ (or $\tilde A$)
modes is {\it intrinsic}, not added from the external with consequent problems of arbitrariness. The intrinsic character of the reference ultimately traces back to  $G_{\mathbf k}$, which, from one side, is algebraically well defined as one of the (three) $SU(1,1)$ generators and, from the other side, as mentioned, is  a term of the Hamiltonian.

\subsection{The semantic value of $q$-bit in the $q$-deformed Hopf coalgebra of QFT}
\label{ssec:bogo1D}

As already observed, the DDF formalism for open system QFT offers the possibility of the foliation of the Hilbert space of the states, namely of   the infinitely many inequivalent representations, i.e. the system huge memory capacity.
At a first sight this could appear as a drawback of the modeling. As a matter of fact, it is not so, because the present scheme also allows the dynamical process of the free energy extremization, leading to the local free energy minimum, which then operates as a ``truth evaluation function'' of the dissipative system ``dynamical" (coalgebraic) semantics. Such a further feature of the doubling procedure plays thus a relevant role as a computational tool and, by itself, may constitute the main motivation for introducing the doubled degrees of freedom. Let us see how this goes.

When dealing with finite temperature QFT, one important task is the computation of thermal averages of some relevant observable, say ${\cal O}$.  This requires the computation of traces. Such a computation is of course of great importance in physics in general and in statistical physics in particular.   Typically, one deals with matrix elements of the type ${\cal O}_{nm} = \langle n|{\cal O}|m\rangle$, with orthonormal states  $\langle n|m\rangle = \delta_{nm} $  in the Fock space and  $H \,| n\rangle = E_{n} \,|n\rangle$. Here $H = \om A^{\dag} A$, $E_{n} = n \,\om $.
The trace $\sum_n {\cal O}_{nn}$ is obtained by multiplying the matrix elements by $\delta_{nm}$ and summing over $n$ and $m$. However, one needs to introduce the $\delta_{nm}$ as an external (to the operator algebra) computational tool, which essentially amounts in picking up ``by hand'' the diagonal elements of the matrix and summing them.
One may instead represent the $\delta_{nm}$ in terms of the doubled tilde-states $|{\tilde n}\rangle$, $\langle {\tilde n}|{\tilde m}\rangle = \delta_{nm} $, with  ${\tilde H}\,|{\tilde n}\rangle = E_{n} \,|{\tilde n}\rangle$ and ${\tilde H} = \om {\tilde A}^{\dag} {\tilde A}$. Then, by using the notation $|n,{\tilde n}\rangle = |n\rangle \times |{\tilde n}\rangle$, since ${\cal O}$ operates only on the non-tilde states, we have
\be \lab{trace}
\langle n,{\tilde n}|{\cal O}|m,{\tilde m}\rangle = \langle n|{\cal O}|m,\rangle\,\langle {\tilde n}|{\tilde m}\rangle = \langle n|{\cal O}|m\rangle\, \delta_{nm} = \langle n|{\cal O}|n\rangle.
\ee
We remark that in the states $|m,{\tilde m}\rangle$ $m$ is an integer number. We have thus a Kronecker delta in Eq.~(\ref{trace}), not a Dirac delta function.
Considering the example of the thermal average $\langle N_{A_{\bf k}} \rangle$ of the number operator $N_{A_{\bf k}}$  we obtain (Takahashi and Umezawa, 1975; Umezawa et al., 1982; Umezawa, 1993)
\be \lab{tracetherm}
\langle N_{A_{\bf k}} \rangle = \frac{Tr[N_{A_{\bf k}} e^{-\beta\,H}]}{Tr[ e^{-\beta\,H}]} = \langle 0(\theta (\beta))| N_{A_{\bf k}}|0(\theta (\beta)) \rangle,
\ee
where the deformation parameter $\theta$ depends on the the temperature $T$ through $\beta = 1/k_{B}T$, with $k_{B}$ the Boltzmann constant, and
$|0(\theta (\beta)) \rangle$ has the form given in Eq.~(\ref{vactheta}). Eq.~(\ref{opnum}) shows that as $\theta$, i.e. the temperature in the present example, changes, the number of condensed modes in the $\theta$-vacuum changes.

This shows that as temperature changes, i.e. $\beta \rar \beta'$, the state spaces (the representations of the CCR) at different temperatures are unitarily inequivalent ones. They are distinct (physically inequivalent) due to their different condensate content for different $\beta(\theta)$ values. Changes of temperature thus are described as trajectories over the space of the unitarily inequivalent representations. Such a space of the representation has been shown to be a K\"ahlerian manifold with a symplectic structure. It has been shown that the trajectories on such a manifold are chaotic classical trajectories (Vitiello, 2004). The K\"ahlerian manifold thus appears as a {\it classical blanket} covering the quantum dynamical evolution in each of the representations.

It is remarkable that in full generality the vacuum state $|0 (\theta) \rangle_{\cal N}$ has a representation in terms of the entropy operator $S_A$. 
By omitting for simplicity the subscript $\cal N$,  $|0 (\theta) \rangle$ can be written indeed as (Umezawa et al., 1982; Umezawa, 1993; Celeghini et al., 1992):
\bea \lab{entrVac} |0 (\theta) \rangle  = \exp{\left ( -
{1\over{2}} S_{A} \right )} |\,{\cal I}\rangle  \, = \exp{\left (
- {1\over{2}} S_{\ti A} \right )} |\,{\cal I}\rangle  ~,
\eea
with the entropy operator
\bea
 S_{A} \equiv  - \sum_{\bf k} \lf( A_{\bf k}^{\dagger} A_{\bf
k} \log \sinh^{2} {\theta}_k - A_{\bf k} A_{\bf k}^{\dagger} \log
\cosh^{2} {\theta}_k \ri) \lab{entr}~,
\eea
and $ |\,{\cal I}\rangle \, \equiv \exp {\left( \sum_{\bf k}
A_{\bf k}^{\dagger} {\ti A}_{\bf k}^{\dagger} \right)} |0\rangle $.
 $S_{{\ti A}}$ has an expression similar to $S_{A}$, with due replacements of
$A$ with ${\ti A}$. We may denote by $S$ either $S_{A}$ or $S_{{\ti A}}$.
In order to get a constraint on the condensate in $|0(\theta)\rangle$,
the free energy  ${\cal F}_{A}$ for the $A$-modes can be introduced:
%
\be\lab{(2.14)} {\cal F}_{A} \equiv \langle 0(\theta)| \Bigl (H_{A}
- {1\over{\beta}} S_{A} \Bigr ) |0(\theta)\rangle~, \ee
where  $H_{A} = \sum_{\bf k} \hbar \omega_{k} A_{\bf
k}^{\dagger} A_{\bf k}$. By minimizing ${\cal F}_{A}$  in each representation,
 ${{\partial {\cal F}_{A}}\over{\partial \theta_k}} = 0 ~,
 \forall ~ {\bf k}$~, we get
\be
\lab{bose}
{\cal N}_{a_{\bf k}} = \sinh^{2}  \theta_k  = {1\over{{\rm
e}^{\beta  E_{k}} - 1}} \, , \ee
with $E_{k} \equiv
\hbar \omega_{k}$, which is the Bose distribution for
$A_{\bf k}$. This holds at each time {\it t} if we allow time dependence of  $\theta$, e.g. in non-equilibrium thermal theories,
$\theta_k = \theta_k (\beta (t))$.

We remark that the above discussion holds for a generic $\theta$ parameter, without any reference to temperature dependence. In that case, one also obtains Eq.~(\ref{bose}), which is recognized to be nothing but the Planck distribution function provided that $\beta$ is taken to be the inverse temperature. Eq.~(\ref{bose}) thus becomes the {\it definition}
of  temperature. Since such a $\theta$ acts as a label of the unitarily inequivalent representations of the CCR, QFT appears then to be intrinsically a thermal field theory (Celeghini et al., 1992). We also recall (cf. Section IV B) that  Thermo Field Dynamics is actually equivalent to the GNS construction in the $C^*$-algebra formalism (Umezawa, 1993; Celeghini et al., 1992).

One may show that time evolution is controlled by the entropy variations (De Filippo and Vitiello, 1977; Celeghini et al., 1992), with consequent introduction of the {\it arrow of time}  (breakdown of time-reversal invariance). By defining  heat as ${dQ={1\over{\beta}} dS}$ and denoting time derivative as  ${\dot{\cal N}}_{a_{\bf k}}$, from $ d {\cal F}_{a} = 0$ we have
\be \lab{dfree} d E_{A} = \sum_{\bf k} \hbar \,\omega_{k} \,
\dot{{\cal N}}_{A_{\bf k}}(t)d t = {1\over{\beta}} d {\cal  S} = d Q ~, \ee
wich relates the change in time $d {\cal N}_{A}$ of condensed particles and heat dissipation $dQ$.

We also recall the relation of the doubling procedure considered above with the
Schmit decomposition process (see e.g. Nielsen and Chuang, 2000; Auletta et al., 2009), where, by denoting by $A$ the modes of the system under study,
the ``doubled" system modes $\tilde A$ are introduced and one works in the composite Hilbert space ${\cal H}_{{A},{\tilde {A}}} \equiv {\cal H}_{A} \otimes {\cal H}_{\tilde {A}}$
with states
$|\Psi_{{A}, \tilde {A}}\rangle  = \sum_n \,\sqrt{w_n} \,|a_n \,{\tilde a}_n \rangle \in {\cal H}_{{A},\tilde {A}}$, $\sum_n w_n = 1$ and $\langle {\tilde a}_m |{\tilde a}_n \rangle  = \delta_{nm}$.  Note that the Schmit decomposition requires the quantum field theory framework in order to obtain physically interesting results (which cannot be gauged away by a unitary transformation). The density matrix for mixed states of the system ${A}$, $\rho^{A} = \sum_n w_{n} |a_n\rangle \langle a_n |$, is obtained by tracing out the system $\tilde {A}$.  In terms of our discussion above, the expression of the state $|\Psi_{{A}, \tilde {A}}\rangle$ has to be compared with
(Umezawa et al., 1982; Umezawa, 1993, Celeghini et al., 1992):
\bsa\label{M3}
  |0 (\theta) \rangle &=& \sum_{n=0}^{+\infty} \sqrt{W_n} \left( |n \rangle
  \otimes |{n} \rangle  \right) ~,
\\ \label{M4}
  W_n &=& \prod_{\bf k}
  \frac{\sinh^{2n_{\bf k}}\theta_k}{\cosh^{2(n_{\bf k}+1)}\theta_k}\,,
\esa
where $n$ denotes the set $\{ {n}_{\bf k} \}$,  $ 0 < W_n
< 1$, $~\sum_{n=0}^{+\infty} W_n = 1$ and the entropy has the familiar form 
$\langle 0(\theta) |S|0(\theta) \rangle =
\sum_{n=0}^{+\infty} W_n \; \log W_n$. 

\subsection{Free energy and entropy for two level system. The generation of the Fibonacci progression.}
\label{ssec:qbit2}

For their obvious relation with boolean logic, it is interesting to consider a two level system, e.g. a two level atom or a qubit, described by the orthonormal basis of two unit (pure state) vectors  $|0\rangle$ and $|1\rangle$, $\langle i| j \rangle = \de_{ij}$, $~i,j = 0,1$. Moreover, we present in this Section the derivation of Fibonacci progression by  recursive processes in the computing framework of QFT.

The state vectors  $|0\rangle$ and $|1\rangle$ are eigenstates of the operator
\be \lab{h} H = \omega_1 |0 \rangle \langle 0| + \omega_2 |1 \rangle \langle 1| ~,
\ee
with eigenvalues $\om_1$ and $\om_2$, respectively: $H |0 \rangle = \om_1 |0 \rangle$ and $H |1 \rangle = \om_2 |1 \rangle$,  $~\om_1 \neq \om_2$. They may denote the eigenvalues of the energy (the frequencies, in natural units $h = 1 = c$) or  another quantum number (charge, spin, etc.). One may then prepare, at some initial time $t_0 = 0$, the superposition of states
\begin{eqnarray}\lab{phi}
|\phi  \rangle & = &
\alpha  \;|0\rangle \;+\;  \beta  \; |1\rangle \,,~
\\ \lab{psi}
|\psi  \rangle & = &
-\beta  \;|0\rangle \;+\; \alpha  \; |1\rangle \,,
\end{eqnarray}
with $|\phi  \rangle$ and $|\psi  \rangle$ not eigenstates of $H$ since $\om_1 \neq \om_2$.
Orthonormality requires that the coefficients $\alpha$ and $\beta$ satisfy the relations $|\alpha |^{2} + |\beta |^{2} = 1$ and $\alpha^{*}\beta-\alpha\beta^{*}=0$. Thus we may set in full generality $\alpha  = e^{i\gamma_{1} } \cos \theta$ and $\beta = e^{i\gamma_{2} } \sin \theta $, with $\gamma_{1}=\gamma_{2}+ n\pi$, $n=0,1,2...$.
For simplicity, one may set $\gamma_1 = 0 = \gamma_2$.
Time evolution of the mixed states $|\phi \rangle$ and $|\psi \rangle$ is then described by
\be \lab{dif}
\left(
     \begin{array}{c}
       |\phi(t)\rangle \\
       [2mm]
       |\psi(t)\rangle \\
     \end{array}
   \right)= e^{-i\omega_{1}t}
   \left(
     \begin{array}{cc}
       \cos\theta & e^{-i(\omega_{2}-\omega_{1})t}\sin\theta \\
       [2mm]
       - e^{-i(\omega_{2}-\omega_{1})t}\sin\theta & \cos\theta \\
     \end{array}
   \right)\left(
            \begin{array}{c}
              |0\rangle \\
              [2mm]
              |1\rangle \\
            \end{array}
          \right) .
\ee
One may show that the generator of time evolution is the free energy operator given by
\begin{eqnarray}\label{FreeEn}
 F \,=\, (H\, -\,\om_{\phi \psi}\,\sigma_{1})
\end{eqnarray}
provided one identifies the term $\om_{\phi \psi}\,\sigma_{1}$ with the entropy term $T S$, with $T$ the temperature and $S$ the entropy.  $\om_{\phi \psi} = \frac{1}{2}
( \omega_{2}- \omega_{1})\;\sin 2\theta \,= \, \langle \psi(t)|\; i\partial_t\;|\phi(t)\rangle = \omega_{\psi \phi}$ and $\sigma_{1}$ is the first of the Pauli matrices:
\bea \nonumber
\sigma_1 =
\frac{1}{2}\left(
  \begin{array}{cc}
    0 & 1 \\
    1 & 0 \\
  \end{array}
\right)~, \quad  \sigma_2 =
\frac{1}{2}\left(
  \begin{array}{cc}
    0 & -i \\
    i & 0 \\
  \end{array}
\right)~, \quad \sigma_3 =
\frac{1}{2}\left(
  \begin{array}{cc}
    1 & 0 \\
    0 & -1 \\
  \end{array}
\right)~.
\eea
We obtain
\begin{eqnarray} \label{TS}
 T S=  \om_{\phi \psi} (|\phi (t) \rangle \langle \psi (t)| + |\psi (t) \rangle \langle \phi (t)|) ~.
\end{eqnarray}
Note that  the ``mixed term'' quantity $\om_{\phi \psi}$  is  responsible for  ``oscillations'' between the states $|\phi (t)\rangle$ and $|\psi (t) \rangle$, occurring since the  $\omega_2 - \omega_1 \neq 0$ and the ``mixing angle" $\theta$ is non-vanishing.

In order to compute the static  and the dynamic entropy for the qubit states $|\phi(t)\rangle$ and $|\psi(t)\rangle$ one needs to double the states by introducing the tilde states  $| \tilde 0 \rangle$ and $| \tilde 1 \rangle$
 \bea\label{fk1}
|0\rangle & \rightarrow & |0\rangle \otimes|\tilde{0}\rangle\,,
\\
\label{fk2}|1\rangle & \rightarrow & |1\rangle \otimes|\tilde{1}\rangle \,.
\eea
The density matrices  are denoted by  $\rho_\xi\,=|\xi(t), {\tilde \xi}(t)\rangle\langle\xi(t), {\tilde \xi}(t)|$ where $\xi\,=\,\phi\,,\psi$ and ${\tilde \xi}\,=\,{\tilde \phi}\,,{\tilde \psi} $.
The reduced density matrix  $\rho^{\cal S}_\phi$ (and similarly for $\rho^{\cal S}_\psi$) is obtained by tracing out the tilde-system ${\tilde {\cal S}}$, and vice-versa. One then may compute the entropies of the mixed state systems (Bruno et al., 2011).

In terms of standard notation $ |0 \rangle = \left(
  \begin{array}{c}
    0 \\
    1 \\
  \end{array}
\right)$ and $|1 \rangle = \left(
  \begin{array}{c}
    1 \\
    0 \\
  \end{array}
\right)$,
we may consider in general a collection of $N$ objects,  labelled by $i = 1,2,...N$ as  $|0 \rangle_i$ and  $|1 \rangle_i$~.
However, for simplicity we omit the index $i$ in our discussion.

By using the matrices $\sigma^{\pm} = \sigma_1 \pm i \,\sigma_2$,
the transitions between the two states $|0 \rangle$ and $|1 \rangle$ can be generated:
\begin{equation} \lab{01}
\sigma^- |1 \rangle =|0 \rangle ~,  \quad ~\sigma^+ |0 \rangle = |1 \rangle ~,  \quad ~\sigma^- |0 \rangle = 0~,~~ \quad ~\sigma^+ |1 \rangle = 0~.
\end{equation}

As usual, the state $|1\rangle$ is said to be the excited state with respect to $|0\rangle$ which is called the `vacuum' or the ground state.
We can show how a ``tree'' can be generated by use of $\sigma^{+}$ and $\sigma^{-}$. We start with $|0 \rangle$.
We do not consider the (trivial) possibility to remain in the initial state $|0\rangle$. In other words, we neglect fluctuations in the ground state, which can be described by $\sigma^- \, \sigma^+ |0 \rangle = 1 |0 \rangle$, i.e. $|0 \rangle \rightarrow |0 \rangle$. In QFT this is obtained by considering the so-called `normal ordering' or `Wick product' of the operators.

The interesting possibility is the one offered by the process leading from  $|0\rangle$ to $|1\rangle$, as in Eq.~(\ref{01}):
\begin{equation} \lab{1}
 |0 \rangle ~~~\rar ~~~\sigma^+ |0 \rangle = |1 \rangle.
\end{equation}
We consider ``single steps'', meaning that we only consider  multiplications by one single matrix, not by a product of $\sigma$'s. Note that $\sigma^+ \, \sigma^+ = 0 = \sigma^- \, \sigma^-$ and that application of $\sigma^+ \sigma^-$ to $|1 \rangle$ is considered to produce a single step, since it is equivalent to the application of the unit matrix $I = \left(\begin{array}{cc}
    1 & 0 \\
    0 & 1 \\
  \end{array}
\right)$ to $|1 \rangle$. In general, for any integer $n$,  $(\sigma^+ \sigma^-)^n \,  |1 \rangle = 1 \times  |1 \rangle$. From Eq.~(\ref{1}), one more single step leads to
\bea \lab{2a}
  &\nearrow& \sigma^- |1 \rangle = |0 \rangle\\  
 |0 \rangle ~~~\rar ~~~\sigma^+ |0 \rangle = |1 \rangle ~~\rar ~~{} \non \\
&\searrow& \sigma^+ \sigma^- |1 \rangle = |1 \rangle \lab{2b}
\eea
%
The `persistence' in the excited state, described by  Eq.~(\ref{2b}), represents a dynamically non-trivial possibility, therefore worth to be considered. One further step leads us to
\bea \lab{12a}
  &\nearrow& \sigma^-  |1 \rangle = |0 \rangle ~~\rar ~\sigma^+ |0 \rangle ~~=~~~ |1 \rangle \\
|0 \rangle ~~\rar ~~\sigma^+ |0 \rangle = |1 \rangle ~~\rar ~~ \non \\
&\searrow& \sigma^+ \sigma^- |1 \rangle = |1 \rangle_{\quad \large{\searrow \sigma^+ \sigma^-  |1 \rangle~= |1 \rangle}}^{\quad \nearrow \sigma^-  |1 \rangle ~~= ~~|0 \rangle} \lab{12b}\\
\non
\eea
\bea
~1  \qquad \qquad  \qquad  1 \qquad  \qquad  \qquad \qquad  \qquad \qquad    2    \qquad  \qquad \qquad  \qquad  ~~ 3 \label{8}
\eea
and so on. Thus we see that at each step, new branching points, or new nodes of the tree, ${}^{\nearrow}_{\searrow} $, are obtained and the tree is generated by recursive $\sigma$ `operations', i.e. by multiplying $|0 \rangle$ and $|1 \rangle$ by the $\sigma$ matrices (the ``SU(2) transformation group'') (Perelomov, 1986).
The `number of the states' in these first steps is in the sequence: $1 ~~1~~ 2 ~~3$, starting with $|0 \rangle$, [one state] , then in equations ~(\ref{1}) [one state],  (\ref{2a}) and  (\ref{2b}) [2 states], and (\ref{12a}) and  (\ref{12b}) [3 states], respectively, (cf. Eq.~(\ref{8})).

Subsequently, from the two $|1 \rangle$'s,  in the next step we have two  $|0 \rangle$'s  and two $|1 \rangle$'s, and from the $|0 \rangle$ we get one single $|1 \rangle$; in total 5 states:  $1 ~~1~~ 2~~ 3~~ 5$. We obtain that  the numbers of states at each step are in the Fibonacci progression ($\{F_n \}, F_{0} \equiv 0$)  with the ones obtained in previous steps. Suppose, indeed, that at the step $F_{p+q}$, one has $p$ states $|0 \rangle$ and $q$ states $|1 \rangle$; in the next step we will have: $(p + q)\,|1 \rangle$ and $q\,|0 \rangle$, $F_{q + (p+q)}$. In the subsequent step: $(p + 2q)\,|1 \rangle$ and $(p + q)\,|0 \rangle$, a total of states $2p + 3q = (q + p + q) + (p + q)$, i.e. the sum of the states in the previous two steps, in agreement with the rule of the Fibonacci  progression construction.

In conclusion, the Fibonacci progression has been generated by iterative algebraic applications of the $\sigma$-matrices, a process which has its  dynamical representation in the Hamiltonian. It is also known (Vitiello, 2009, 2012, 2014) the relation between fractal-like self-similar systems and the Fibonacci progression, which is ubiquitous in inorganic and organic systems, and also in linguistics (Piattelli-Palmarini and Vitiello, 2016) and brain functioning (Freeman and Vitiello, 2006, 2008; Vitiello, 2009). The isomorphism between coherent states in DDF QFT, and self-similar fractal structure is also known (Vitiello, 2012, 2014).

\section{Toward a computational interpretation of a Thermal QFT}

\subsection*{Quantum vacuum as an ``infinite state black-box machine" }
\label{ssect:qv}
The evidence that our coalgebraic modeling of thermal QFT can naturally generate Fibonacci's progressions forces us to interpret it as an \textit{universal quantum computing structure}, in the framework of ``Universal Coalgebra" as general theory of systems, interpreted as ``labelled state transition systems" (Rutten 2000). On the other hand, the computational significance of the DDF principle of thermal QFT can be fully appreciated when we consider that the \textit{q}-deformed Hopf coalgebras and algebras, constitute in thermal QFT\ two \textit{dually equivalent} categories for the contravariant application of a same functor \textit{T}, that is, the Bogoliubov transform (see Sect. \ref{ssec:bogol} and Appendix \ref{dual_alg_coalg}).

This dual equivalence, indeed, not only removes the main obstacle for a complete application of a \textit{coalgebraic semantics } in quantum computations, that is the lack of \textit{contravariance} in these constructions (Abramsky, 2013), but also it allows us to apply
to thermal QFT the coalgebraic construction of the ``infinite state black-box machine". This implies the non-insignificant specification that the ``states" here concerned are as many ``phase coherence domains" of the QV foliation, including a quantum open system and its thermal bath. The consequences for a theoretical biology based on a thermal QFT, the cognitive neurosciences included, are easy to imagine. Let us discuss briefly this conclusive point.

Generally, in the coalgebraic interpretation of computing systems, a \textit{state transition system }(STS) is an abstract machine, consisting of a (finite/infinite) set of states, and a (finite/infinite) set of transitions, which may be \textit{labelled} or \textit{unlabelled}, with labels chosen from a set (Rutten, 2000; Venema, 2007). Finite STS distinguishes a special "start" state, and a set of special "final" states.
Formally, a STS, is a pair $(S, \rightarrow)$, where $S$ is a set of states, and \(((\rightarrow)\subseteq S\times S)\) is a transition binary relation over $S$. If $p, q$ belong to $S$ and $(p, q)$ belongs to $(\rightarrow)$, then $(p \rightarrow q)$, i.e., there is a transition from $p$ to $q$.
Moreover, a labelled STS (LTS) is a triple $(S, \Lambda, \rightarrow)$, where $\Lambda$ is a set of labels, and \(((\rightarrow)\subseteq S \times \Lambda \times S)\) is a ternary relation of labelled transitions. If, $p, q$ belong to $S$, $\alpha$ belongs to $\Lambda$, and $(p, \alpha, q)$ belongs to $(\rightarrow)$, then $\left( {p\mathop  \to \limits^\alpha  q} \right)$. Label set $\Lambda$ is thus a set of maps $\{\alpha: p\mapsto q\}$ characterizing a class of LTS.

In computer science, a label generally corresponds to a program, so that given the natural interpretation of a program as a Boolean algebra, it is possible to interpret its semantics, as far as defined on the states of the (abstract) machine in which the program is implemented, as the corresponding coalgebra, with the evident advantage that such semantics can apply to functional programming over infinite sets, such as \textit{streams}  (Rutten, 2000).

Let us consider now  a dynamic system represented as a LTS on a functorial coalgebra for some functor $\Omega$ that admits a \textit{final coalgebra} (Venema, 2007).

We recall the definition of a final coalgebra for a functor $\Omega$.

A functor $\Omega: \mathbf{C}\rightarrow \mathbf{C}$ is said to \textit{admit a final coalgebra} if the category \textbf{Coalg}$(\Omega)$ (see Appendix \ref{dual_alg_coalg}) has a final object, that is, a coalgebra $\mathbb Z$ such that from every coalgebra $\mathbb A$ in \textbf{Coalg}$(\Omega)$ there exists a unique homomorphism $!_\mathbb A: \mathbb A \rightarrow \mathbb Z$.

We remark that by the notion of final coalgebra it is possible to give a formal definition of the  notion in quantum physics of  \textit{observational equivalence} expressing the fact  that two states of the same system or of two systems are indistinguishable provided their observables (i.e., their behaviours, in the following denoted as \textit{beh}) are the same.

More generally, following (Venema, 2007),
let $\mathbb S = \langle \mathbb S, \sigma \rangle$ and $\mathbb S' = \langle \mathbb S', \sigma' \rangle$ two systems for the set functor $\Omega$. The states $s \in \mathbb S$ and $s' \in \mathbb S'$ are \textit{observationally equivalent}, i.e., $\mathbb S, s\equiv_\Omega \mathbb S', s'$, if there exists an $\Omega$-system $\mathbb X \langle X, \xi \rangle$ and homomorphisms $f: \mathbb S \rightarrow \mathbb X$ and $f': \mathbb S' \rightarrow \mathbb X'$, such that $f(s) = f'(s')$. In case $\Omega$ admits a final coalgebra $\mathbb Z$,  $\mathbb S, s\equiv_\Omega \mathbb S', s'$ iff $!_\mathbb S (s) =\ !_{\mathbb S'} (s')$.

One of the most interesting applications of these notions  for the theory of dynamic systems as STS's, both for computer science and QFT, is the possible coalgebraic formalization of a particular infinite state dynamic system, the so-called ``black-box machine''. Following (Venema, 2007), given that the simplest characterization of a system is that of a $C$-colored ($C$-valued) set, i.e. a pair $\langle S,\gamma : S\rightarrow C\rangle$ such that for whichever state the system can only display the color (value) associated to the current state, and halt after doing so, a ``black- box machine'' is a machine whose internal state are invisible to an external observer. That is, it is a system prompted to display a given value or color from $C$, and to move to the next state. What is observable are thus only the ``behaviors'' of this machine.

Formally,
a black-box machine $\mathbb M = \langle M, \mu \rangle$ is an abstract machine that can be modeled as a coalgebra $\mu: M\rightarrow C\times M,$ with $\pi _0(\mu(s)) \in C$ denoting the current value (color) of the machine state, and $\pi _1(\mu(s))\in M$ representing the machine's next internal state, and where $\pi_0: C\times M\rightarrow C$ and $\pi_1: C\times M\rightarrow M$ are the projections functions.

Starting from the state $x_0$ the machine makes a transition $\mu (x_0) = (c_0, x_1)$, and continues with $\mu (x_1) = (c_1, x_2),\:\mu (x_2) = (c_2, x_3), ...$ . Because the states $x_0, x_1, x_2,$... are internal to the machine, the only observable part of this dynamics is the infinite sequence or \textit{stream}: $beh(x_0) = (c_0, c_1, c_2, ...…) \in C^\omega$ of values in the data set $C$.

The collection $C^\omega$ of all the infinite ``words'' over $C$ forms itself a system for the functor $C \times\mathcal{I}$,  where $\mathcal{I}$ is the set of all identity functions $Id_x$ identifying univocally each object $x$ in $C$. Simply endow the set $C^\omega$ with the transition structure splitting an infinite stream $u = c_0, c_1, c_2,...…$ into its two \textit{coordinates}: \textit{head} $h(u) = c_0$ and its \textit{tail} or remainder:  $t(u) = c_1, c_2, c_3,...…$ . Putting $\gamma (u) = \langle h(u), t(u)\rangle$, one easily proves that the behaviour map: $x \mapsto beh(x)$ is the unique homomorphism from $\mathbb M$ to its coalgebra $\langle C^\omega, \gamma \rangle$. This shows that $\langle C^\omega, \gamma \rangle$ is the final object in the category \textbf{Coalg}$(\mathbf{C}^\omega,\mathbf{\gamma} )$.

In order to make explicit the relation between the notion of black-box machine in computer science and thermal QFT, it is useful to recall that in the manifold of the unitarily inequivalent representation $\{|0(\theta)\rangle; \forall \theta \}$, the quantum vacua (QV) $|0(\theta)\rangle$ differ among themselves due to a different coherent condensate content of NG modes, referred to as ``long range quantum entanglement modes" at the beginning of Sect. III. Transitions ({\it phase transitions}) through such QV (through the respective Hilbert spaces), is induced by the generator $G_\mathbf{k}$ of the Bogoliubov transformations (cf. Section IV A). In terms of the previous construction \(\mathbb{M}\), $G_\mathbf{k}$ corresponds to the transition structure \(\gamma\). This generator produces the univocal labeling of the QV $|0(\theta)\rangle_{\mathcal N}$ by the order parameter value  $\cal N$-set $\equiv \{ {\cal N}_{A_{\bf k}}(\theta), {\cal N}_{A_{\bf k}}(\theta) = {\cal N}_{{\tilde A}_{\bf k}}(\theta), \forall {\bf k}  \}$.
The set \(\mathcal{N}\) corresponds in the construction \(\mathbb{M}\) to the set $\mathcal{I}$ of all the $\{Id_x\}$ indexing univocally the $q(\theta)$-\textit{foliation }of the QV, 
obtained by inverting the Bogoliubov transformations for all the quantum numbers \textbf{k} and acting  as the contravariant functor $T$. This means that the collection $T^\omega$ of all the infinite representations of the CCR's over $T$ forms itself a system for the functor $T \times\mathcal{I}$, i.e., the so-called ``diagonal functor" (see Appendix \ref{dual} ) of our construction.

The trajectories through the manifold $\{|0(\theta)\rangle; \forall \theta \}$ can be described, as already mentioned, as classical chaotic trajectories in the space of the representations $\{|0(\theta)\rangle; \forall \theta \}$ (cf. Section III D). These trajectories appear as macroscopic (classical) manifestations of the microscopic dynamics ruling the changes at the quantum level of the QV condensates, which thus remain hidden to the observer by the ``classical blanket'', indeed (cf. Section IV D).
An open question is whether, in such a formalism, we are giving a possible effective realization of Feynmann proposal of mimicking the computational process in terms of a path integrals QM, where the most probable path belongs to a bundle of unobservable paths.

We thus see that the QFT structure has built in the  ``transition structure $\gamma$'' of a black-box machine. We can state then that under the action of the functor $T$, the \textit{dual equivalence} between the categories of the $q$-deformed Hopf coalgebras and algebras is established,
i.e. $\mathbf{qHCoalg (T)}\rightleftharpoons\mathbf{qHAlg (T^{op})}$.
We have, indeed, the unique transition structure $\gamma$ characterizing the category of the $q$-deformed Hopf coalgebras and of their ``mirrored'' algebras, through the contravariant application of the functor \textit{T} (See Appendix \ref{dual_alg_coalg}).

\section{Conclusions }
In this paper, using CT formalism,  we gave a first survey of the possibility and of the fruitfulness of interpreting the coalgebraic modeling of thermal QFT as TFD of quantum dissipative systems in the framework of the ``Universal Coalgebra", namely as a general theory of dynamic and computation systems as ``labelled state transition systems" (Rutten, 2000). We showed the fruitfulness of this interpretation from the standpoint: (1) of a topological formalization  of the coalgebraic modeling of thermal QFT systems, the biological systems included; and (2) of the consequent possibility of developing topological quantum computing models of thermal QFT systems.
 The theoretical connective of these two topological interpretations lies in the evidence that the \textit{q-}deformed Hopf coalgebras and the doubled Hilbert spaces of thermal QFT share the same topological spaces of \textit{C*}-algebras, and of the Stone-spaces, on which the topological interpretations of QFT and of quantum computing are respectively based (see Sect. \ref{ssec:Cstar} and Appendix \ref{dual_alg_coalg}).

    As to (1), the Universal Coalgebra offers the proper formal framework in which to justify the two main results of the coalgebraic modeling of dissipative quantum systems in thermal QFT. That is: (a) the possibility, via the DDF principle of determining \textit{dynamically} the orthonomal basis of the Hilbert space  of the quantum system (see Sect. \ref{ssec:doubl}); (b) the connected possibility of a \textit{dynamic} determination of the diagonal elements of the matrix for the trace computation of thermal averages of a given observable in finite temperature QFT. At the same time, this doubling of the matrix elements, avoids the Dirac delta function. Indeed, the states on which the observable operates are represented by integer numbers, so that we have a Kronecher delta, and not a Dirac delta function (see Sect. \ref{ssec:bogo1D}).

As to (2), the Universal Coalgebra offers the proper formal framework in which to justify the two main results demonstrating the possibility of topological quantum computations based on thermal QFT systems (see Sect. \ref{ssec:qbit2}).  That is: (a) the semantic value of the qubit in thermal QFT systems, because the doubling of the states allows to interpret the connected measure of the free-energy minimum as an evaluation function for the associated \((\top/\negmedspace\perp)\) quantum Boolean operator; (b) the evidence that the iterative applications of the \(\sigma\)-matrices associated with this semantic qubit generates a Fibonacci progression which has its \textit{dynamical} representation in the Hamiltonian of the system.

The possibility of interpreting such results in the Universal Coalgebra framework is finally demonstrated in the Appendix \ref{dual_alg_coalg} , and in Sect. \ref{ssect:qv}. In fact, the DDF principle characterizing thermal QFT has its most natural topological formalization in the dual equivalence between the categories of \(\mathbf{qHCoalg}\) and of \(\mathbf{qHAlg}\) of thermal QFT, via the ``Bogoliubov construction". That is, by the contravariant vectorial mapping on Hilbert spaces related to the Bogoliubov transform, \textit{T*}, by using in a contravariant way the endofunctor \textit{T} characterizing the category \(\mathbf{qHCoalg}\). The consequent interpretation of the category \textbf{qHCoalg} as a \textit{final coalgebra} for the category of ``diagonal functors" \((T\times\mathcal{I})\) given in Sect. \ref{ssect:qv} allows us to state that such a class of topological quantum computing architectures is endowed with an \textit{infinite} memory capacity (QV foliation), even though made dynamically available on a \textit{finitary} (neither ``finitistic", nor ``infinitistic") basis, in the sense defined by Abramsky as ``limits/colimits of finite successions" (Abramsky, 2005).

On the other hand, the evidence recalled in the Introduction that the QV-foliation in QFT has been successful applied to solve dynamically the capacity problem of the long-term memories (deep beliefs) in the living brain, interpreted as a ``dissipative brain" entangled with  their environments, finds in this interpretation its theoretical justification.

Anyway, this interpretation opens the way  toward a topological quantum computing based on ordinary phases of matter, before all the photonic ones, by using, according to a coalgebraic logic, a technology of optical quantum computing now well-established, even though till now used within the conventional quantum statistical mechanics theoretical framework. However, also in this case, it is displaying intriguing capabilities of computing non-Turing statistical functions (Carolan et al., 2015).  Concretely, all this means that, for implementing ``long range quantum entanglements", in our case including also the streams, with their ``topological orderings" depending on the non-commutative products  of \textit{q-}deformed Hopf coalgebras, we could use (suitable arrays of) classical Mach-Zender optical interferometers. They will work over electromagnetic front-waves at the environment temperatures,  instead of using gallium arsenide interferometers, working over statistical waves of non-abelian anyons, at \(\approx0^{\circ}\)K temperature.
 These intriguing theoretical and practical perspectives justify further researches in this new direction.

\appendix

\section*{Appendices}

For the reader convenience, we summarize here  some key notions of coalgebraic logic used in the paper. Particularly, we concentrated ourselves on the duality coalgebras-algebras because of the contravariant application of a same functor. On the other hand, ``duality" is a notion widely applied in logic (think at the fundamental De Morgan's laws), in mathematics (think at a function and at its inverse), and in physics (think at a function $f(x)$ and at its Fourier transform $\hat f(\xi)$) (Atiyah, 2007).
Nevertheless, its role in logic and mathematics can be fully appreciated only in the context of CT formalism. For this aim in the following three appendixes we introduce some basic notions of CT, then we concentrate on some functorial dual constructions in CT, for arriving finally at the illustration of the duality between functorial algebras-coalgebras, as distinct from the axiomatic justification of duality of these structures that is the normal case in topological QFT.

\section{Some basic notions of Category Theory}
\label{sect:app}

In this brief summary we refer in particular to (Abramsky \& Tzevelekos, 2011) for defining some basic notions of CT used in this paper (see also (Awodey, 2010) for a more complete treatment of CT formalism).

CT is a universal language in logic and mathematics that is in many senses more general than set theory. The principal difference as to set theory is that in CT the primitives are: 1) \textit{morphisms} or \textit{arrows}, $f, g$, – intended as a generalization of notions such as ``function", ``operator", ``map", etc., –, and 2) the \textit{compositions of arrows}, $f \circ g$. In this way, even set elements in CT are to be considered as domain-codomains of morphisms, and, more generally any ``object" \textit{x} in CT corresponds to the domain of a reflexive morphism $I_x$, i.e., to an ``identity" relationship.

Therefore, a category is any mathematical structure with structure-preserving morphisms:

\vspace{0,3cm}

{\bf Definition} (Definition of category). A \textit{category }\textbf{C} consists of:

\begin{itemize}
\item
A collection, \textbf{Ob(C)},  of  \textit{objects} denoted by $A,B,C...$; \item
A collection, \textbf{Ar(C)}, of \textit{arrows} (morphisms) denoted  by \textit{f, g, h, ...};
\item
\textit{Mappings} \textbf{dom}, \textbf{cod}: $\textbf{Ar(C)}\rar \textbf{Ob(C)}$ assigning to each arrow $f$ its domain $\textbf{dom}(f)$ and its codomain $\textbf{cod}(f)$, respectively;
\item
For any triple of objects $A, B, C$, a \textit{composition map}:  \({A\mathop \to \limits^f B\mathop \to \limits^g C}\), written as $g\circ f$ (or sometimes, $f;g$);
\item
For any object \textit{A}, an identity arrow $id_A: A\rar A$.
\end{itemize}
The above must satisfy the following axioms:
\begin{displaymath}
h\circ (g\circ f)= (h\circ g)\circ f,\qquad\  f\circ id_A = f =id_B \circ f.
\end{displaymath}
Whenever the domains and codomains of the arrows match appropriately, so that the compositions are well defined.

\vspace{0,2cm}

{\bf Remark 1.} Examples of categories of some fundamental algebraic structures are thus: \textbf{Set} (sets and functions); \textbf{Grp} (groups and homomorphisms); $\textbf{Vect}_{k}$ (vector spaces over a field $k$, and linear maps); \textbf{Pos }(partially ordered sets and monotone functions); \textbf{Top} (topological spaces and continuous functions), etc.

\vspace{0,2cm}

\textbf{Remark 2. Pos} is fundamental in logic and mathematics. Indeed, partial ordering is a structure of ordering relations $(\leq)$ among sets satisfying simultaneously:
1) \textit{Reflexivity} $(x \leq x)$; 2) \textit{Antisymmetry} $(x \leq y \wedge\ y\leq x \Rightarrow x = y)$; and 3) \textit{Transitivity} $x\leq y \wedge y\leq z \Rightarrow x\leq z$.
The structure of "total ordering" of sets, and the relative category \textbf{Tos} of totally ordered sets, satisfies \textit{Antisymmetry} and \textit{Transitivity} but instead of \textit{Reflexivity} it satisfies the ordering property of \textit{Totality} $(x\leq y \vee y\leq z)$. That is, for all sets an ordering relations is defined. Nevertheless, the category \textbf{Tos} lacks in an "object" as to \textbf{Pos}, because the ordering relation $(\leq)$ is no longer an object in it, since it does not satisfy any longer \textit{Reflexivity} like in \textbf{Pos}. Therefore, \textbf{Tos} is a subcategory of \textbf{Pos}. In fact, fundamental posets are the real number set $(\mathbb{R},\leq)$, and the power set \(\mathcal{P}\) of a given set \textit{X} $(\mathcal{P}(X),\subseteq)$.

 \vspace{0,2cm}

Another fundamental notion of CT is the notion of \textit{functor} \textit{F}, that is a "morphism between categories" sending all the objects, arrows, and compositions from a category \textbf{C} into another \textbf{D}, i.e., \textit{F}: \textbf{C} \(\rightarrow\) \textbf{D}, so to justify a homomorphism, or bijective mapping, between the categories \textbf{C}  and \textbf{D}.  Of course, if the homomorphism can be reversed we have an \textit{isomorphism.} Moreover, for each category \textbf{C}, there exists an \textit{endofunctor, E,} mapping a category onto itself: i.e., \textit{E}: \textbf{C} \(\rightarrow\) \textbf{C}.

Finally, the application of a functor \textit{F} between the categories \textbf{C}  and \textbf{D} can be \textit{covariant}, if it preserves all the objects, and the directions of the morphisms and of the orders of compositions between the two categories, so that the two categories are \textit{equivalent} as to the functor \textit{F}, \textbf{C} \(\equiv_{F}\) \textbf{D}. On the contrary, the application of a functor \textit{G} between the categories \textbf{C}  and \textbf{D} is \textit{contravariant} if it preserves all the objects, but reversing all the directions of the morphisms between objects (i.e., from $A\rightarrow B$, to $GB\rightarrow GA$), and the orders of the compositions between morphisms (i.e., from $f\circ g$ to $Gg\circ Gf$). In this case, the target category of the functor is the \textit{opposite} one as to the source category. That is, a functor \textit{G} is contravariant, if $G: \mathbf{C}\rightarrow \mathbf{D}^{\mathbf{op}}$. In this case, the two categories are \textit{dually equivalent} as to the functor \textit{G}, i.e., \(\mathbf{C}\rightleftharpoons_{G}\mathbf{D}^{\mathbf{o\mathbf{p}}}\).
 Of course, it is possible also to have a contravariant application of an endofunctor $E: \textbf{C}\rar \textbf {C}$. In this case, the contravariant application of this functor on all the morphisms and compositions, links the category \textbf{C} to its opposite \(\mathbf{C}^{\mathbf{op}}\), i.e., \(E^{op}:\mathbf{C}\rar \mathbf{C}^{\mathbf{op}}\). In other terms, there exists a \textit{homomorphism up to isomorphism} between $\mathbf{C}$ and $\mathbf{C}^{\mathbf{op}}$ through $E^{op}$.

\section{Categorical dualities in CT logic and in computer science}
\label{dual}
The notion of \textit{opposite category}, according to which a category \textbf{C}  is dual as to its opposite \(\mathbf{C}^{\mathbf{o\mathbf{p}}}\), leads us immediately to the fundamental \textit{principle of duality} in logic and specifically in \textit{semantics}, according to which a statement \textit{S} is true in/about \textbf{C} if and only if (\textit{iff}) its dual \(S^{op}\), obtained from \textit{S} by reversing all the arrows is true in/about \(\mathbf{C}^{\mathbf{o\mathbf{p}}}\). That is, \textit{S} and \(S^{op}\) are \textit{logically dually equivalent }, i.e., \(S\leftrightarrows\ S^{op}\). Note that this is significantly different from the ordinary equivalence between two statements \(S, S'\) defined in/about the same category \textbf{C}, i.e., \(S\leftrightarrow S'\).

What is important to emphasize for our aims is that in computational set theoretic semantics the dual category of \(\mathbf{Set^{\mathbf{op}}}\) is more important than \textbf{Set}, given that, for instance a generic conditional in logic ``if...then", e.g., ``for all \textit{x}, if \textit{x} is a horse, then \textit{x} is a mammalian", is true iff the ``mammalian set" includes the ``horse set" with all its subsets. Therefore, the semantics of a given statement is set theoretically defined on the powerset \(\mathcal{P}(X)\) of a given set \textit{X}. From this the categorical definition of the \textit{covariant }powerset functor \(\mathcal{P}:\mathbf{Set}\rightarrow\mathbf{Set}\) derives:

\vspace{0,2cm}

{\bf Definition} (Categorical definition of the covariant powerset functor). The powerset functor \(\mathcal{P}\) is a covariant endofunctor \(\mathbf{Set}\rightarrow \mathbf{Set}\), mapping each set \textit{X} in its powerset \(\mathcal{P}(X)\) and sending each function \(f: X\rightarrow Y\) to the map \textit{S} that sends \(U\subseteq X\) to its image \(f(U)\subseteq Y\), that is:
\begin{eqnarray}
\begin{array}{c}
X\mapsto\mathcal{P}(X). \\
(f:X\rar Y)\mapsto\mathcal{P}(f):= S\mapsto\{f(x)| x \in S\}\\
\end{array}
\end{eqnarray}

\vspace{0,2cm}

Conversely, the definition of the contravariant power set functor \(\mathcal{P}^{op}:\mathbf{Set^{op}}\rightarrow\mathbf{Set}\) is as following:

\vspace{0,2cm}

{\bf Definition} (Categorical definition of the contravariant power set functor).
The contravariant power set functor \(\mathcal{P}^{op}\) is a contravariant endofunctor \(\mathbf{Set^{op}}\rightarrow\mathbf{Set}\) that preserves all the objects, and sends each function \(f:X\rightarrow Y\) to the map \textit{T} sending \(V\subseteq Y\) to its inverse image \(f^{-1}(V)\subseteq X\). Therefore:

\begin{eqnarray}
\begin{array}{c}
\mathcal{P}^{op}(X) :=\mathcal{P}(X). \\
\mathcal{P}^{op}(f:X\rar Y):\,\,\,\mathcal{P}(Y)\longrightarrow\mathcal{P}(X):= T\mapsto\{x\in X|f(x)\in T\}\\
\end{array}
\end{eqnarray}

\vspace{0,2cm}

Another useful example for us, because used, both in the GNS-construction for \textit{C*-}aòlgebras , and in the Vietoris construction for Boolean Algebras, is the dual space functor $(\_)^{*}$ on vector spaces \textit{V} defined on a field \textit{k}:
\begin{equation}
(\_)^{*}: \mathbf{Vect}^{op}_{k}\rar \mathbf{Vect}_{k}:= V\mapsto\ V^{*}
\end{equation}

The logical significance of these two last constructions, the contravariant powerset functor, and the dual space functor for vector spaces is evident. ``These are both examples of the same idea: send an object \textit{A} to the functions from \textit{A} into some fixed object" (Abramsky-Tzevelekos, 2011). E.g., we can think at the powerset in terms of the characteristic function of a subset: \(\mathcal{P}(X)=2^X\). Roughly speaking, given the ``humans" as a subset of ``mammalians" in the real world, we can ``induce" the predicates expressing truthfully ``being human" in our representational world, without attaining at any ``higher order" semantics.

Moreover, in CT other useful categorical dual constructions can be significantly formalized that we cannot define here, but that have an immediate significance for us because both the formalism of quantum physics and of quantum computation are plenty of exemplifications of their usage. For instance, the notions of "\textit{left" and "right adjoints}" of functions and operators, the notions of "\textit{universality}" and "\textit{co-universality}",  of "\textit{limits}" and "\textit{colimits}" interpreted, respectively, as "\textit{final}" and "\textit{initial}" objects of two categories related by a third category of "\textit{indexing functors}", \(\mathcal{I}\), so to grant the mapping, via a \textit{diagonal functor}, of all the objects and morphisms of one category into the other. We saw the usefulness of such a last construction in the formalization of the ``black-box infinite state machine" (see Sect. \ref{ssect:qv}).

Practically, all the objects and the operations that are usefully formalized in set theory, and then in calculus and set-theoretic logic,– included the "\textit{exponentiation}" operation for forming function spaces, and the consequent "evaluation function" over function domains –, can be usefully formalized also in CT, with a significant difference, however. Instead of considering objects and operations for what they "\textit{are}" as it is in set theory, in CT we are considering them for what they "\textit{do}" (Abramsky and Tzevelekos, 2011).
Just what we need for interpreting QFT in the framework of TFD!
\section{Duality between functorial coalgebras and algebras}
\label{dual_alg_coalg}

We apply now the notion of duality also to the categorical pair \textit{coalgebras-coalgebras,} for the contravariant application of a functor \(\Omega\) characterizing them. We know that, \textit{per se}, both the category of Hopf algebras, \textbf{HAlg} -- which are  \textit{bi-algebras} with an \textit{antipode}, that is, a linear vectorial mapping between commutative co-products and products and co-units and units --, and the category of Hilbert spaces, \textbf{Hil,} are \textit{self-dual}. Nevertheless, self-duality is instead non valid in the usage of both these structures in thermal QFT,  where the thermal deformation parameter \textit{q} breaks the symmetry coalgebra-algebra of Hopf bi-algebras, making \textit{non-commutative} the co-products of the coalgebra, and, correspondingly, the  \textit{composite} Hilbert space includes the system states $\mathcal H_A$, and their entangled environment states $\mathcal{H}_{\tilde A}$,
according to the tensorial product $\mathcal{H}_{A,\tilde A} \equiv \mathcal{H}_{A} \otimes \mathcal{H}_{\tilde A}$.

We start with the definition of the category of coalgebras for a generic functor $\Omega$ (Venema, 2007), because coalgebras in CT logic have a semantic primacy as to their dually related algebras, just like \(\mathbf{Set^{op}}\) as to \(\mathbf{Set}\), and just like the thermal bath as to an open system in thermal QFT.
Therefore, following (Venema, 2007):
\vspace{0,3cm}

{\bf Definition} (Coalgebra for a functor $\Omega$).
Given an endofunctor $\Omega$ on a category \textbf{C}, an $\Omega$-coalgebra is a pair \(\mathbb{A}=\langle A,\alpha\rangle\), where \textit{A} is an object of \textbf{C} called the \textit{carrier} of $\mathbb A$, and $\alpha:A\rightarrow \Omega A$ is an arrow in \textbf{C}, called the \textit{transition map} of $\mathbb A$. In case $\Omega$ is an endofunctor on \textbf{Set}, $\Omega$-coalgebras
may also be called $\Omega$-\textit{systems}; a \textit{pointed $\Omega$-system}\footnote{This terminology is related with the representation of sets like as many ``accessible pointed graphs" (apg). In each apg the root node or ``vertex" is (decorated with) the set, the other nodes are (decorated with) its subsets, and the oriented edges are inclusion relations.} is a triple \(\langle A,\alpha,a \rangle\) such that \(\langle A,a \rangle\) is an $\Omega$-system and \textit{a} is a state in $\mathbb A$, that is an element of \textit{A}.
\label{def:coalg_funct}

\vspace{0,3cm}
{\bf Definition} (Homomorphism between $\Omega$-coalgebras).
Let \(\mathbb{A}=\langle A,\alpha\rangle\) and \(\mathbb{A'}=(A',\alpha')  \) be two coalgebras for the functor $\Omega: \mathbf{C}\rightarrow\mathbf{C}$. Then a homomorphism from \textit{A} to \textit{A'} is an arrow $f : A \rightarrow A'$ for which the
following diagram commutes:
\begin{displaymath}
\begin{CD}
A @>f>> A' \\
@V{\alpha}VV @V{\alpha'}VV \\
\Omega A @>\Omega f>> \Omega A \\
\end{CD}
\end{displaymath}
It is easy to demonstrate that the collection of coalgebra homomorphisms contains all identity arrows and it is closed under arrow composition. Therefore, the $\Omega$-coalgebras form a \textit{category}.

\vspace{0,3cm}

{\bf Definition} (Definition of an $\Omega$-coalgebra  category).
For any functor $\Omega:\mathbf{C}\rightarrow\mathbf{C}$ let $\textbf{Coalg}(\Omega)$ denote the category of $\Omega$-coalgebras as objects and the corresponding homomorphisms as arrows. The category \textbf{C} is called the \textit{base category} of $\textbf{Coalg}(\Omega)$.
\label{def:omega_coalg_cat}

Let us pass now to the definition of the category of algebras for a generic functor $\Omega$. For this let us start from recalling that an \textit{algebra} over a signature $\Omega$ is a set \textit{A} with an $\Omega$-indexed collection \(\{f^\mathbb{A}|A^{ar(f)}\rightarrow A\}\)\footnote{We recall that with the \textit{arity} (\textit{ar}) or \textit{rank} of a function/operation we denote the number of arguments/operands to which the function/operation applies.} of operations. These operations may be combined into a single map \(\alpha:\sum_{f \in \Omega} A^{ar(f)}\rightarrow A\), where \(\sum_{f \in \Omega} A^{ar(f)}\) denotes the \textit{coproducts} (or \textit{sum} or \textit{disjoint union}) of the sets \(\{A^{ar(f)}| f\in \Omega\}\). From this it is possible to verify that a map \(g: A\rightarrow A'\) is an \textit{algebraic homomorphism} iff  the following diagrams commutes:
\begin{displaymath}
\begin{CD}
A @>f>> A' \\
@A{\alpha}AA @A{\alpha'}AA \\
\Omega A @>\Omega f>> \Omega A' \\
\end{CD}
\end{displaymath}
where we see now the signature $\Omega$ as the polynomial set \textit{functor} \(\sum_{f \in \Omega} \mathcal{I}^{ar(f)}\). That is $\Omega$ operates also on functions between sets, and not only on sets. This suggests the following definition, generalizing to algebras the previous two ones:

\vspace{0,3cm}
{\bf Definition} (Definition of an $\Omega$-algebra category).
Given an endofunctor $\Omega$ on a category \textbf{C}, an $\Omega$-algebra is a pair \(\mathbb{A}=\langle A,\alpha\rangle\) where \(\alpha: \Omega A \rightarrow A\) is an arrow in \textbf{C}. A homomorphism from an $\Omega$-algebra $\mathbb A$ to an $\Omega$-algebra $\mathbb A'$ is an arrow \(f: A\rightarrow A\) such that \(f \circ \alpha = \alpha' \circ \Omega f \). The category so \textit{induced} is denoted as \textbf{Alg}$(\Omega).$
\label{def:omega_alg_cat}

\vspace{0,3cm}

It is now evident the similarities between algebras and coalgebras for a given functor $\Omega$. Effectively the name ``coalgebra" is justified from the fact that a coalgebra \(\mathbb{C}=\langle C,\gamma: C\rar \Omega C\rangle\) over a base category $\textbf{C}$ can also be seen as an algebra in the \textit{opposite} category $\mathbf{C}^{op}$. I.e.:
\begin{equation}
\mathbf{Coalg}(\Omega)=(\mathbf{Alg}(\Omega^{op}))^{op}.
\end{equation}
In this categorical framework, it is therefore easy to define formally both the dual equivalence between the category of the coalgebras defined on the Stone-spaces, \textbf{SCoalg}, and the category of Boolean algebras, \textbf{BAlg}, by a contravariant vectorial mapping between them, and the dual equivalence between the category of \textit{q-}deformed Hopf coalgebras, \(\mathbf{qHCoalg}\) and algebras, \(\mathbf{qHAlg}\) of thermal QFT. Indeed, evidently, both \(\mathbf{qHCoalg}\) and the category \textbf{Stone} of  Stone-spaces share the same topology of the \textit{C*}-algebras. I.e., these topological spaces are \textit{compact} (in the topological sense), \textit{totally disconnected} or \textit{pointwise} (in a topological space, only the one-point sets (the unitary and the empty sets) are connected\footnote{For sake of completeness, we recall that the empty set is excluded from the posets defined on Stone spaces of the Stone representation theorem for Boolean algebras. Indeed, such posets are as many ``ultrafilters" defined on the powerset of a given set.}), \textit{Hausdorff} (its points have disjoint neighbourhoods) spaces.

Generally, (Venema, 2007) every endofunctor \(\Omega\) on \textbf{Stone} induces an endofunctor \(\Omega^*:= (\cdot)_*\circ\Omega\circ(\cdot)^*\) on \textbf{BAlg}. Therefore, it is an immediate consequence of (C1) that the categories \(\mathbf{Coalg(\Omega)}\) and \(\mathbf{Alg(\Omega^*)}\) are dually equivalent:
\begin{equation}
\mathbf{Coalg(\Omega)\rightleftharpoons Alg(\Omega^*)}
\end{equation}
An exemplification of this fact is the well-known dual equivalence between \textbf{SCoalg} and \textbf{BAlg} via the so-called ``Vietoris construction", i.e., by using the endofunctor \(\mathcal{V^*}\) on the Vietoris space sharing the same topology of \textbf{Stone}:
\begin{equation}
\mathbf{SCoalg(\mathcal{V})\rightleftharpoons\mathbf{BAlg(\mathcal{V}^*)}}
\end{equation}

The further exemplification of (C2), introduced in this paper and mainly in Sect \ref{ssec:bogol}, is the dual equivalence between \(\mathbf{qHCoalg}\) and \(\mathbf{qHAlg}\) of thermal QFT, via the ``Bogoliubov construction". That is, by the contravariant vectorial mapping on Hilbert spaces related to the Bogoliubov transform, \textit{T*}, by using in a contravariant way the endofunctor \textit{T} characterizing the category \(\mathbf{qHCoalg:}\) \begin{equation}
\mathbf{qHCoalg(T)\rightleftharpoons\mathbf{qHAlg(T^*)}}
\end{equation}

\vspace{0,4cm}

\subsubsection*{\bf Acknowledgements}
The Ministero dell'Istruzione, dell'Universit\`a e della Ricerca (MIUR) and the Istituto Nazionale di Fisica Nucleare (INFN) are acknowledged
for partial financial support.

\vspace{0,2cm}

\subsubsection*{\bf Declaration of interest}
All authors declare that they do not have any actual or potential conflict of interest including any financial,
personal or other relationships with other people or organizations within three years of beginning the
submitted work that could inappropriately influence, or be perceived to influence, their work.

\newpage

{\bf \large Bibliography}\\


Abramsky, S., 2005. A Cook's Tour of the Finitary Non-Well-Founded Sets (original lecture: 1988), in: Artemov, S. et al. (Eds.), Essays in honor of Dov Gabbay. Vol. I, Imperial College Pubblications, London, 2005, pp. 1 - 18.\\

Abramsky, S., Tzevelekos, N., 2011. Introduction to categories and categorical logic, in: Coecke, B., (Ed.), New structures for physics. Lecture Notes in Physics, vol. 813,  Springer , Berlin-New York, 2011, pp. 3 - 94.\\

Abramsky, S., 2013.  Coalgebras, Chu Spaces, and Representations of Physical Systems. J. Phil. Log. 42, 551 - 74.\\

Aczel, P., 1988. Non-Wellfounded Sets. CLSI Lecture Notes, vol.14.\\

Aczel, P., Mendler, N.P., 1989. A Final Coalgebra Theorem", in: Category Theory and Computer Science, Lecture Notes in Computer Science, vol. 389. Springer, London, UK.\\

Atiyah, M. F., 2007. Duality in mathematics and physics, in: Lecture notes from the Institut de Matematica de la Universitat de Barcelona (IMUB), $<http://fme.upc.edu/ca/arxius/butlleti-digital/riemann/071218_conferencia_atiyah-d_article.pdf> $ \\

Auletta, G.,  Fortunato, M., Parisi, G., 2009.  Quantum Mechanics. Cambridge University Press, Cambridge.  \\

Awodey, S., 2010. Category Theory. Second Edition (Oxford Logic Guides 52), Oxford UP, Oxford, UK.\\

Basti, G., 2013. A change of paradigm in cognitive neurosciences Comment on: Dissipation of `dark energy' by cortex in knowledge retrieval, by Capolupo, Freeman and Vitiello. Physics of life reviews, 5, 97 - 98.\\

Basti, G., 2017. The quantum field theory (QFT) dual paradigm in fundamental physics and the semantic information content and measure in cognitive sciences, in: Dodig-Crnkovic, G., Giovagnoli, R.(Eds.), Representation of Reality: Humans, Animals and Machine, Springer Verlag , Berlin New York, 2017, in Press.\\

Blasone, M., Srivastava, Y.N., Vitiello, G., Widom, A., 1998. Phase Coherence in Quantum Brownian Motion.
Annals of Physics 267, 61 - 74.\\

Blasone, M.; Jizba, P.; Vitiello, G., 2011. Quantum Field Theory and its Macroscopic Manifestations. Imperial College Press, London.\\

Bratteli, O., Robinson, D. W., 1979.  Operator Algebras and Quantum Statistical Mechanics. Springer Verlag, Berlin.\\

Bruno, A., Capolupo, A., Kak, S., Raimondo, G., Vitiello, G., 2011.
Gauge theory and two level systems.
Modern Physics Letters B  25, 1661 - 1670.\\

Capolupo, A., Freeman, W.J., Vitiello, G., 2013.
Dissipation of dark energy by cortex in knowledge retrieval. Physics of Life Reviews, 5, 90 - 96.\\

Carolan, J. et al., 2015. Universal linear optics. Science 349, 711 - 16.\\

Celeghini, E., Rasetti, M., Vitiello, G., 1992. Quantum dissipation. Annals of Physics  215,  156 - 170.\\

Celeghini, E., De Martino, S., De Siena, S., Iorio, A., Rasetti, M., Vitiello, G., 1998.
Thermo field dynamics and quantum algebras.
Physics Letters A 244,  455 - 461.\\

Connes, A., Rovelli, C., 1994. Von Neumann algebra automorphisms and time-thermodynamics relation in general covariant quantum theories. Classical and Quantum Gravity 11, 2899 - 918.\\

De Filippo, S.,  Vitiello,  G., 1977. Vacuum structure for unstable particles. Lettere al
Nuovo Cimento 19, 92 - 98.\\

Emch, G.G., 1972. Algebraic Methods in Statistical Mechanics and Quantum Field Theory. Wiley-Interscience, New York. \\

Freedman, M. H. et al., 2002. Topological quantum computation. Bull. of the Amer. Mathem. Soc. (New Series), 40, 31 - 38.\\

Freeman, W.J., Vitiello, G., 2006.
Nonlinear brain dynamics as macroscopic manifestation of underlying many-body field dynamics. Physics of Life Reviews 3,  93 - 118.\\

Freeman, W.J., Vitiello, G., 2008. Brain dynamics, dissipation and spontaneous
breakdown of symmetry. Journal of Physics A: Mathematical and Theoretical 41,
304042.\\

Goldstone, J.,  Salam, A.,  Weinberg, S., 1962. Broken symmetries. Physics Review
127, 965–970.\\

Haag, R., Hugenholtz, N.M., Winnink, M., 1967.
On the equilibrium states in quantum statistical mechanics.
Comm. Math. Phys. 5,  215 - 236.\\

Holland, S.,  Longo, R., 2016. Non Equilibrium Thermodynamics in Conformal Field Theory. in arXiv:1605.01581v1 [hep-th]
<http://www.arxiv.org>\\

Kubo, R., 1957. Statistical-Mechanical Theory of Irreversible Processes. I. General Theory and Simple Applications to Magnetic and Conduction Problems.
J. Phys. Soc. Japan, 12, 570 - 586.\\

Landsman, N.P., 2011. Lecture notes on operator algebras, in Institute for Mathematics, Astrophysics, and Particle Physics. Radboud University Nijmegen <http://www.math.ru.nl/~landsman/OA2011.html>

Martin, P.C., Schwinger, J., 1959. Theory of Many-Particle Systems. I.
Phys. Rev. 115, 1342 – 1373
doi:https://doi.org/10.1103/PhysRev.115.1342\\

Nayak C., et al., 2008. Non-Abelian anyons and topological quantum computation. Rev. Mod. Phys. 80, 1083.\\

Nelson, E., 1986. Predicative Arithmetic (Mathematical Notes 32), Princeton University Press, Princeton, NJ.\\

Nielsen, M.A., Chuang, I.L., 2000. Quantum Computation and Quantum Information. Cambridge University Press,  Cambridge. \\

Ojima, I., 1981. Gauge fields at finite temperatures - "Thermo field dynamics" and the KMS condition and their extension to gauge theories.
Annals of Physics 137, 1 - 32.\\

Perelomov, A. M., 1986. Generalized coherent states and their applications. Springer Verlag, Berlin.\\

Perrone, A.L., 1995. A formal Scheme to Avoid Undecidable Problems. Applications to Chaotic Dynamics, Lecture Notes in Computer Science 888,
9 - 48.\\

Piattelli-Palmarini, M., Vitiello, G., 2016. Linguistics and Some Aspects of Its Underlying
Dynamics. Biolinguistics 9, 96 - 115.\\

Rutten, J.J.M.,  2000. Universal coalgebra: a theory of systems. Theoretical computer science 249, 3 - 80.\\

Sangiorgi, D., Rutten, J.J.M., (Eds.), 2012. Advanced topics in bisimulation and coinduction, Cambridge UP , New York.\\

Stone, M.H., 1936. The theory of representation for Boolean algebras. Trans. of the Am. Math. Soc. 40, 37 - 111.\\

Takahashi, Y. Umezawa, H., 1975. Thermo field dynamics. Collect. Phenom. 2, 55 - 80. Reprinted in Int. J. Mod. Phys. B  10, 1755 - 1805 (1996).\\

Takesaki, M., 1970.
Tomita's Theory of Modular Hilbert Algebras and its Applications, Springer, Berlin/Heidelberg/New York. \\

Umezawa, H., Matsumoto, H., Tachiki, M., 1982. Thermo Field Dynamics and condensed states. North-Holland, Amsterdam.\\

Umezawa, H., 1993. Advanced field theory: Micro, macro, and thermal physics. American Institute of Physics, New York.\\

Venema, Y., 2007. Algebras and co-algebras, in: Blackburn P., van Benthem, F.A.K., Wolter, F., (Eds.),  Handbook of modal logic. Elsevier, Amsterdam,  pp. 331 - 426.\\

Vitiello, G., 1995. Dissipation and memory capacity in the quantum brain model.
Int. J.  Mod. Phys. B 9, 973 – 989.\\

Vitiello, G., 2001. My Double Unveiled. John Benjamins, Amsterdam.\\

Vitiello, G., 2004. Classical chaotic trajectories in quantum field theory. Int. J.  Mod. Phys. B 18, 785 – 792.\\

Vitiello, G., 2009.
Coherent states, fractals and brain waves.
New Math. and Natural Computation 5, 245 - 264.\\

Vitiello, G., 2012.
Fractals, coherent states and self-similarity induced noncommutative geometry.
Phys. Lett. A 376, 2527 - 2532.\\

Vitiello G., 2014. On the Isomorphism between Dissipative Systems,
Fractal Self-Similarity and Electrodynamics.
Toward an Integrated Vision of Nature. Systems 2014 2, 203 - 216.
doi:10.3390/systems2020203\\

Vitiello, G., 2015.
The use of many-body physics and thermodynamics to describe the dynamics of rhythmic generators in sensory cortices engaged in memory and learning.
Current Opinion in Neurobiology 31,  7 - 12.\\

von Keyserlingk, C. W.,   Simon, S. H.,  Rosenow, B., 2015. Enhanced Bulk-Edge Coulomb Coupling in Fractional Fabry-Perot Interferometers. Phys. Rev. Lett. 115, 126807.\\

von Neumann, J., 1955. Mathematical foundations of Quantum Mechanics. Priceton University Press, Princeton.\\

Willett, R.L.  et al., 2013. Magnetic field-tuned Aharonov-Bohm oscillations and evidence for non-Abelian anyons at v = 5/2. Phys. Rev. Lett. 111, 186401.\\


\end{document}